\documentclass[a4paper,12pt]{article}
\usepackage{amsfonts,amsthm,amsmath,amssymb,graphicx}
\begin{document}

\newtheorem{definition}{Definition}[section]
\newcommand{\be}{\begin{equation}}
\newcommand{\ee}{\end{equation}}
\newcommand{\bea}{\begin{eqnarray}}
\newcommand{\eea}{\end{eqnarray}}
\newcommand{\LE}{\left[}
\newcommand{\R}{\right]}
\newcommand{\nn}{\nonumber}
\newcommand{\Tr}{\text{Tr}}
\newcommand{\N}{\mathcal{N}}
\newcommand{\G}{\Gamma}
\newcommand{\vf}{\varphi}
\newcommand{\LL}{\mathcal{L}}
\newcommand{\HH}{\mathcal{H}}
\newcommand{\arctanh}{\text{arctanh}}
\newcommand{\up}{\uparrow}
\newcommand{\down}{\downarrow}
\newcommand{\ket}[1]{\left| #1 \right>}
\newcommand{\bra}[1]{\left< #1 \right|}
\newcommand{\ketbra}[1]{\left|#1\right>\left<#1\right|}
\newcommand{\rd}{\partial}
\newcommand{\de}{\partial}
\newcommand{\ba}{\begin{eqnarray}}
\newcommand{\ea}{\end{eqnarray}}
\newcommand{\db}{\bar{\partial}}
\newcommand{\we}{\wedge}
\newcommand{\ca}{\mathcal}
\newcommand{\lr}{\leftrightarrow}
\newcommand{\f}{\frac}
\newcommand{\s}{\sqrt}
\newcommand{\vp}{\varphi}
\newcommand{\hvp}{\hat{\varphi}}
\newcommand{\tvp}{\tilde{\varphi}}
\newcommand{\tp}{\tilde{\phi}}
\newcommand{\ti}{\tilde}
\newcommand{\ap}{\alpha}
\newcommand{\pr}{\propto}
\newcommand{\mb}{\mathbf}
\newcommand{\ddd}{\cdot\cdot\cdot}
\newcommand{\no}{\nonumber \\}
\newcommand{\la}{\langle}
\newcommand{\lb}{\rangle}
\newcommand{\ep}{\epsilon}
 \def\we{\wedge}
 \def\lr{\leftrightarrow}
 \def\f {\frac}
 \def\ti{\tilde}
 \def\ap{\alpha}
 \def\pr{\propto}
 \def\mb{\mathbf}
 \def\ddd{\cdot\cdot\cdot}
 \def\no{\nonumber \\}
 \def\la{\langle}
 \def\lb{\rangle}
 \def\ep{\epsilon}

\begin{titlepage}
\thispagestyle{empty}

\begin{flushright}
YITP-14-42\\
IPMU-14-0123\\
\end{flushright}

\vspace{.4cm}
\begin{center}
\noindent{\Large \textbf{Entanglement of Local Operators in large N CFTs}}\\
\vspace{2cm}

Pawe{\l} Caputa $^{a}$,
Masahiro Nozaki $^{a}$,
 and
Tadashi Takayanagi $^{a,b}\,\footnote{email: \{pawel.caputa, mnozaki, takayana\} AT yukawa.kyoto-u.ac.jp}$
\vspace{1cm}

{\it
 $^{a}$Yukawa Institute for Theoretical Physics (YITP),\\
Kyoto University, Kyoto 606-8502, Japan\\
$^{b}$Kavli Institute for the Physics and Mathematics of the Universe (Kavli IPMU),\\
University of Tokyo, Kashiwa, Chiba 277-8582, Japan\\
}

\vskip 2em
\end{center}

\vspace{.5cm}
\begin{abstract}
In this paper, we study Renyi and von-Neumann entanglement entropy of excited states created by local operators in large $N$ (or large central charge) CFTs. First we point that a naive large $N$ expansion can break down for the von-Neumann entanglement entropy, while it does not for the Renyi entanglement entropy. This happens even for the excited states in free Yang-Mills theories. Next, we analyze strongly coupled large $N$ CFTs from both field theoretic and holographic viewpoints. We find that the Renyi entanglement entropy of the excited state produced by a local operator, grows logarithmically under its time evolution and its coefficient is proportional to the conformal dimension of the local operator.
\end{abstract}

\end{titlepage}

\tableofcontents

\section{Introduction}

The entanglement entropy (EE) and its generalization called the Renyi entanglement entropy (REE) have been actively studied recently as a useful tool to understand quantum many-body systems, quantum field theories as well as gravitational theories. In quantum systems, they play a role of good quantum order parameters which characterize the degrees of freedom hidden in various ground states. In particular, they also reveal essential properties in conformal field theories (CFTs) \cite{HLW,CC} and topological field theories \cite{wen}. One main advantage of considering entanglement entropy is that it offers us a geometrical interpretation of many basic properties of quantum systems. Indeed, this becomes manifest in the holographic entanglement entropy (HEE) \cite{RT} when we apply the AdS/CFT correspondence \cite{Maldacena} or more generally holographic principle \cite{Hol}.

To define the entanglement entropy, we divide the total space into two subregions $A$ and $B$ so that the total Hilbert space is factorized ${\cal H}={\cal H}_A\otimes {\cal H}_B$. We define the reduced density matrix $\rho_A$ with respect to the subsystem $A$ from the original density matrix $\rho$ by tracing out ${\cal H}_B$ such that $\rho_A=\mbox{Tr}_{{\cal H}_B}\rho$.
 The $n$-th Renyi entanglement entropy (REE) for subsystem $A$ is defined by
\be
S^{(n)}_A=\f{\log\mbox{Tr}[(\rho_A)^n]}{1-n}. \label{san}
\ee
The (von-Neumann) entanglement entropy (EE) is defined as the $n\to 1$ limit:
\be
S^{(1)}_A=-\mbox{Tr}[\rho_A\log \rho_A].
\ee

So far, most works on entanglement entropy have focused on Renyi entanglement entropies for ground states. The main purpose of this paper is to offer a general picture of entanglement entropy for a simple class of excited states which are obtained by acting local operators on the vacua in CFTs. The studies for such local operator excitations were initiated in slightly different setup in \cite{UAM} and then \cite{Nozaki:2014hna}. We will pursuit the latter approach in this paper. We will be interested in the difference of $S^{(n)}_A$ between the excited state and the ground state, denoted by $\Delta S^{(n)}_A$. We choose the subsystem $A$ to be a half of the total space. Recently, analytical calculations of $\Delta S^{(n)}_A$ have been performed for massless free scalar fields in \cite{Nozaki:2014hna,Nozaki:2014uaa} and rational two dimensional CFTs \cite{He:2014mwa}. In the latter paper, it was proven that $\Delta S^{(n)}_A$ coincides with the quantity called quantum dimension of the primary operator. The present paper will start with a brief review of these previous results so that readers can understand the whole picture.

However, these previous results were obtained in CFTs which do not have tractable holographic duals via AdS/CFT.  Therefore the main purpose of this work is to consider large $N$ gauge theories or large $c$ CFTs as they can possibly have holographic duals. We will perform both field theoretic and holographic analysis and explain how they are consistent with each other. In particular, we would like to understand how the large $N$ limit affects the calculations of entanglement entropy. We will find that the behavior of von-Neumann entanglement entropy $(n=1)$ is rather different from that of Renyi entanglement entropy $(n>1)$ with respect to the large $N$ scaling in our examples.
Note that such a difference does not usually occur for ground states of CFTs, where both von-Neumann and Renyi entropies scale as $c\sim N^2$. Furthermore in strongly coupled and large $N$ CFTs, our holographic and field theoretic calculations suggest that more complicated subtlety appears in the $n\to 1$ limit. At the same time we will find a new behavior of the time evolution of Renyi entanglement entropy which is peculiar to strongly coupled large $N$ CFTs.

To motivate our work further, note that a universal relation similar to the first law of thermodynamics, called the first law-like relation \cite{BNTU,BCHM,WKPW,Firstlaw}, has been known for an excited state produced by a small perturbation.\footnote{Refer also to \cite{Einstein} for recent studies on the connections between Einstein equation and fundamental constraints satisfied by the entanglement entropy (such as the first law-like relation or its generalizations).} However, we cannot apply this to our problem as we choose the subsystem $A$ to be a half of the total space and thus its size is much larger than the length scale of the excitation.

  We would also like to mention that there are other classes of excited states which have been studied well, called quantum quenches \cite{cag,cal}. Refer to e.g.\cite{QHEE,NNT,localq} for their holographic dual calculations. Since quantum quenches are triggered by sudden changes of Hamiltonian, they generate infinitely many operators acted on the vacua. Therefore we can say that our local operator excitations are more mild and elementary, though some of computation techniques are common to both. One more interesting class will be excited states with energy fluxes. They are holographically dual to plane wave geometries and have been studied in  \cite{planewave}.

  This paper is organized as follows: In section 2 we will review the calculations of (Renyi) entanglement entropy of local operator excited states with a few pedagogical examples. In addition, we will present a brief analysis of energy density of our excited states.
  In section 3, we will study (Renyi) entanglement entropy of local operator excited states in a large $N$ free Yang-Mills theory. In section 4, we will study the same entanglement entropy in two dimensional CFTs with a large central charge. In addition we will give a review of previous results in rational 2d CFTs. In section 5, we will perform a holographic computation of the Renyi entanglement entropy of the local operator excited states. In section 6, we briefly mention a holographic result of von-Neumann entanglement entropy for these excited states. In section 7 we will summarize our results. In appendix A, we present the details of calculations of $2n$ point functions in large $c$ two dimensional CFTs. In appendix B, we give some details of geodesic calculations, used for the holographic analysis of Renyi entropy. In appendix C, we present details of direct computation of Renyi entropy of excited states in the free Yang-Mills theory.

\section{Renyi Entanglement Entropy of Local Operator Excited States}

Here we give a brief review of replica method calculations of (Renyi) entanglement entropy of
local operator excited states following \cite{Nozaki:2014hna,Nozaki:2014uaa} (see also \cite{He:2014mwa,Mos}). For a basic construction of replica method in quantum field theories refer to e.g.\cite{CC}.

Consider a QFT in $d$ dimensional flat spacetime $R^{1,d-1}$, whose time and $d-1$ dimensional space are denoted by $t$ and $x^i\ (i=1,2,\ddd ,d-1)$. We focus on an excited state $|\Psi_O\lb$ defined by acting a local operator $O(\tau,x^i)$ on the vacuum state $|0\lb$ at a point $x^i$ and time $t=0$ in a given QFT: $|\Psi_O\lb=O(0,x^i)|0\lb$.
Its time evolution under the Hamiltonian $H$ is described by
\be
|\Psi_O(t)\lb=\s{\mathcal N}\cdot e^{-iHt}e^{-\ep H}O(0,x^i)|0\lb, \label{lesq}
\ee
where $\mathcal{N}$ is a normalization and an infinitesimal parameter $\ep$ is the regularization of the ultraviolet behavior of local operator, which, as we will see later in subsection 2.3, makes the energy of this excited state finite. 

Now we choose the location of insertion of the local operator $O$ to be
\be
x^1=-l(<0),\ \ (x^2,x^3,\ddd,x^{d-1})={\bf x},
\ee
at the time $t=0$. The final result does not depend on the choice of the vector ${\bf x}$ owing to the translational invariance.

We would like to compute the (Renyi) entanglement entropy  $S^{(n)}_A$ defined by (\ref{san}) in the replica method (see e.g.\cite{CC}), performing the Euclidean continuation $\tau=it$. To define the (Renyi) entanglement entropy, we choose the subsystem $A$ to be the half space $x_1>0$. It is useful to introduce a complex coordinate as follows:
\begin{equation}
w =x^1+i  \tau, \ \ \ \bar{w}=x^1-i\tau.
\end{equation}
The density matrix at real time $t$ for the locally excited states (\ref{lesq}) is written as
\ba
\rho(t)&=&{\mathcal N}\cdot e^{-iHt}e^{-\ep H}O(0,x^i)|0\lb\la 0|{O}^{\dagger}(0,x^i)e^{-\ep H}e^{iHt} \no
&=& {\mathcal N}\cdot O(w_2,\bar{w_2}, {\bf x})|0\lb\la 0|
{O}^{\dagger}(w_1,\bar{w}_1, {\bf x}),  \label{densany}
\ea
where the normalization factor ${\mathcal N}$ is determined by the condition Tr$\rho(t)=1$.
Also here we defined
\ba
&& w_1=i(\epsilon -it)-l, \ \ w_2 = -i(\epsilon+it)-l,   \label{wco} \\
&& \bar{w}_1=-i(\ep-it)-l,\ \ \bar{w}_2=i(\epsilon+it)-l.  \label{wcor}
\ea
Throughout the computation we treat $\ep\pm it$ as purely real numbers until the end of calculations as in \cite{cag,NNT,Nozaki:2014uaa}.

Now, if we define $\Delta S^{(n)}_A$ by subtracting the ground state result from $S^{(n)}_A$,
then $\Delta S^{(n)}_A$ is computed as
\begin{equation}\label{sre}
\Delta S_A^{(n)} = \frac{1}{1-n}\log{\left(\frac{\Tr \rho_A^n}{\Tr (\rho^{(0)}_{A})^n}\right)},
\end{equation}
where $\rho^{(0)}_A$ is the ground state reduced density matrix.

By thinking of the path-integral description, as usual in the replica method, $\Delta S_A^{(n)}$ can be expressed in terms of the correlations functions of operator that excites the state \cite{Nozaki:2014hna,Nozaki:2014uaa}:
\begin{equation}\label{gre}
\begin{split}
&\Delta S_A^{(n)} = \frac{1}{1-n}\left(\log{\frac{Z_n}{Z^{(0)}_n}}-n\log{\frac{Z_1}{Z^{(0)}_1}}\right) \\
&=\frac{1}{1-n}\left( \log{\left \langle O^{\dagger}(w_1, \bar{w}_1, {\bf x})O(w_2, \bar{w}_2, {\bf x}’)\cdots O(w_{2n}, \bar{w}_{2n}, {\bf x}’) \right\rangle_{\Sigma_n}}\right) \\
&~~~~~~~~~~~~~~~~~~~~~~~~~~~~~~~~~~~~~-\frac{n}{1-n} \left(\log{\left\langle O^{\dagger}(w_1, \bar{w}_1, {\bf x})O(w_2, \bar{w}_2, {\bf x}’)\right\rangle_{\Sigma_1}}\right),
\end{split}
\end{equation}
where $Z_n$ and $Z^{(0)}_n$ corresponds to the partition functions on $\Sigma_n$ such that
$\Tr \rho_A^n=\f{Z_n}{(Z_1)^n}$ and $\Tr (\rho^{(0)}_{A})^n=\f{Z^{(0)}_n}{(Z^{(0)}_1)^n}$.
The $n$-sheeted $d$ dimensional space $\Sigma_n$ is given by the Fig.1,
\begin{figure}[t]\label{rep}
  \centering
  \includegraphics[width=8cm]{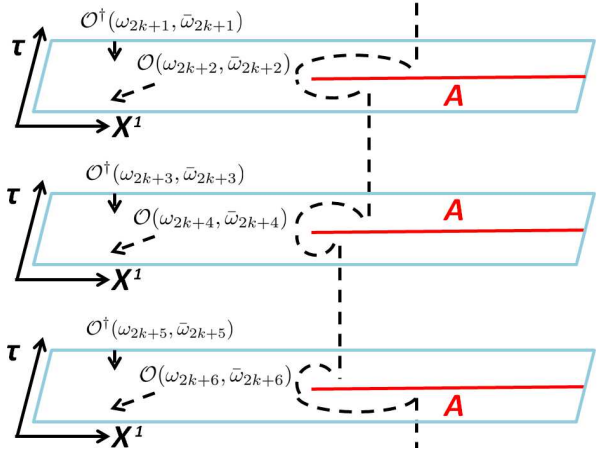}
  \caption{The $n$-sheeted geometry $\Sigma_n$, constructed by gluing the subsystem A on a sheet to another subsystem A on other sheet.}
\end{figure}
where
$2n$ operators $O$ and $O^{\dagger}$ are located periodically as
\be
w_{2k+1}=e^{2\pi ik}\cdot w_1,\ \ \ w_{2k+2}=e^{2\pi ik}\cdot w_2.
\ee
The geometry of $\Sigma_n$ is obtained by pasting $n$ copies of the flat space $R^{d}(=\Sigma_1)$ such that the angle at the origin $w=\bar{w}=0$ is extended
from $2\pi$ to $2\pi n$. In other words, the metric of $\Sigma_n$ is simply given by
introducing the polar coordinate as $w=\rho e^{i\theta}$:
\be
ds^2=d\rho^2+\rho^2 d\theta^2+\sum_{i=2}^{d-1} (dx_i)^2,
\ee
with $0\leq \theta \leq 2\pi n$ and $0\leq \rho<\infty$.

\subsection{Free Massless Scalar Field in Four Dimensions}

As an example, consider a free massless scalar field $\phi$ in four dimension \cite{Nozaki:2014hna,Nozaki:2014uaa}.
We would like to briefly review the real time evolution of the Renyi entropy for the states excited by acting $\phi$ on the ground state (see \cite{Nozaki:2014uaa} for more details).
The density matrix for such state is defined by
\begin{equation}
\rho(t) = {\mathcal N}\cdot \phi(w_2, \bar{w}_2, {\bf x}’)\left|0\right\rangle\left\langle 0\right|\phi(w_1, \bar{w}_1, {\bf x}). \label{localfm}
\end{equation}
We need to compute a $2n$ point function of $\phi$ on $\Sigma_n$ and its two point function on the flat plane as in the formula (\ref{gre}).
The propagator on $\Sigma_n$ \cite{Nozaki:2014hna,Nozaki:2014uaa} is given by
\begin{equation}
G_n(r, s, \theta, \theta', {\bf x}, {\bf x'}) =\frac{1}{4n\pi^2 r s (a-a^{-1})}\frac{a^{\frac{1}{n}}-a^{-\frac{1}{n}}}{a^{\frac{1}{n}}
+a^{-\frac{1}{n}}-2\cos{\frac{\theta-\theta'}{n}}},  \label{gfnwq}
\end{equation}
where $a$ is defined by
\begin{equation}
\frac{a}{1+a^2}=\frac{r s}{\left|{\bf  x}- {\bf x'}\right|^2+r^2+s^2}.
\end{equation}
Then we perform an analytic continuation (\ref{wco}) and (\ref{wcor}) to study the real time evolution.

As shown in \cite{Nozaki:2014uaa}, only specific propagators on $\Sigma_n$ can contribute to the $2n$ point function of $\phi$ on $\Sigma_n$ in the $\epsilon \rightarrow 0$ limit. More precisely, only two disconnected Feynman diagrams can contribute to the the $2n$ point function of $\phi$ on $\Sigma_n$ in this limit (refer to Fig.3 for the two diagrams). The propagator on $\Sigma_1$ is simply given by
\begin{equation}
\left\langle \phi(w_1, \bar{w}_1, {\bf x}) \phi(w_2, \bar{w}_2, {\bf x}’)\right \rangle_{\Sigma_1} =\frac{1}{16\pi^2 \epsilon^2}.
\end{equation}

 In this way, we can evaluate the Renyi entanglement entropy for the locally excited states (\ref{localfm}) by the replica method. Below we would like to show final results for $n=2$ and $n=3$:
 \begin{eqnarray}
\Delta S^{(2)}_A=\left\{ \begin{array}{ll}
0 & 0< t < l, \\
\log{2}-\log{\left[1+\frac{l^2}{t^2}\right]} & t \ge l ,\\
\end{array} \right.
\end{eqnarray}
and
\begin{eqnarray}
\Delta S^{(3)}_A=\left\{ \begin{array}{ll}
0 & 0< t < l, \\
-\frac{1}{2} \log{\left[\frac{1}{4}+\frac{3 l^2}{4 t^2}\right]} & t \ge l .\\
\end{array} \right.
\end{eqnarray}
The time evolution of these Renyi entropies are plotted in Fig.2.

In the $\epsilon \rightarrow 0$ limit and at late time, the final value of the second and third Renyi entanglement entropy is given by
\begin{equation}
\Delta S^{(2)}_A \to \log{2},\ \ \ \Delta S^{(3) }_A \to \log{2}.
\end{equation}
Moreover we can prove the same result $\Delta S^{(n)}_A$ for any $n$ \cite{Nozaki:2014uaa}.
This entropy $\log 2$ coincides with that of an EPR state (maximally entangled state of
two spins). Indeed, we can explain this in terms of entangled pair.
We decompose $\phi$ to the left moving mode and the right moving mode,
\begin{equation}
\phi =\phi_L +\phi_R.  \label{lrdecf}
\end{equation}
After taking into account the normalization, the locally excited state is expressed by
\begin{equation}
\left| \Psi_\phi \right \rangle =\frac{1}{\sqrt{2}}\phi_L\left|0\right\rangle_L \otimes \left|0\right\rangle_R + \frac{1}{\sqrt{2}} \left|0\right\rangle_L\otimes\phi_R \left|0\right\rangle_R. \label{eprfr}
\end{equation}
Thus this expression (\ref{eprfr}) manifestly shows that the excited state at late time becomes an an EPR state \cite{Nozaki:2014hna,Nozaki:2014uaa}. This is because the division of the total space into two halves $A$ and $B$ implies that, in the late time limit $t\to\infty$, two parts of the entangled pair propagate in the opposite (left and right) directions.

More generally, if we consider an excited state obtained by acting $\phi^k$ on the ground state, it can be written as
\begin{equation}
\left|\Psi_{\phi^k}\right \rangle = \frac{1}{2^{\frac{k}{2}}}\sum^{k}_{m=0}\sqrt{ _k C_m}\left|m\right\rangle_A \otimes \left|k-m\right \rangle_B,
\end{equation}
where $_k C_m=\f{k!}{m!(k-m)!}$. From this expression, it is straightforward to evaluate the finite value of $\Delta S^{(n)}_A$ at late time. In \cite{Nozaki:2014uaa}, it was proved that this evaluation precisely matches that obtained from the replica calculation (\ref{gre}).

\begin{figure}\label{l10n2n3}	
  \centering
  \includegraphics[width=8cm]{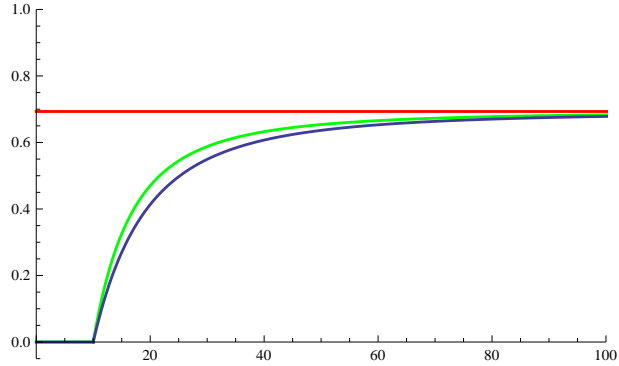}
  \caption{The plots of $\Delta S^{(2)}_A$ (green) and $\Delta S^{(3)}_A$ (blue) as functions of
  the time $t$  in the $\epsilon \rightarrow 0$ limit (we chose $l=10$). The red horizontal corresponds to the late time value $\log{2}$.}
\end{figure}

\subsection{2d CFTs and Free Scalar Example}

As another simple example, let us consider two dimensional CFTs (2d CFTs) and study the replica calculations in explicit examples.
 We focus on an excited state which is defined by acting a primary operator $O$ on the vacuum $|0\lb$ in a given 2d CFT. The (chiral) conformal dimension of this operator is defined as $\Delta_O$. We employ the Euclidean formulation and introduce the complex coordinate
\be
(w,\bar{w})=(x+i\tau,x-i\tau),
\ee
 on $R^2$ such that $\tau$ and $x$ are the Euclidean time and the space, respectively. We insert the primary operator $O$ at $x=-l<0$ and consider its real time-evolution from time $0$ to $t$ under the Hamiltonian $H$. This process is described by the following density matrix:
\ba
\rho(t)&=&{\mathcal N}\cdot e^{-iHt}e^{-\ep H}O(-l)|0\lb\la 0|{O}^{\dagger}(-l)e^{-\ep H}e^{iHt} \no
&=& {\mathcal N}\cdot O(w_2,\bar{w_2})|0\lb\la 0|
{O}^{\dagger}(w_1,\bar{w}_1), \label{dmtd}
\ea
where the normalization factor ${\mathcal N}$ is again determined by the normalization condition Tr$\rho(t)=1$. Note that (\ref{dmtd}) is simply the two dimensional version of (\ref{densany}) and
we defined $w_1$ and $w_2$ as in (\ref{wco}) and (\ref{wcor}). An infinitesimal positive parameter $\ep$ is again an ultraviolet regularization and we treat
$\ep\pm it$ as purely real numbers until the end of calculations as in \cite{cag,NNT,Nozaki:2014uaa}.

To calculate $\Delta S^{(n)}_A$, we employ the replica method in the path-integral formalism
which we explained in the previous subsection.
We choose the subsystem $A$ to be an interval $0\leq x\leq L$ at $\tau=0$.
The replica method calculation involves the partition function on  a $n$-sheeted Riemann surface $\Sigma_n$ with $2n$ operators $O$ and its conjugate.
In the end, we find that $\Delta S_A^{(n)}$ can be computed as
\ba
&&\Delta S_A^{(n)} \no
&&\!=\!\f{1}{1-n} \Biggl[\!\log{\left\langle{O}^{\dagger}_a(w_l,\bar{w}_1)O(w_2,\bar{w}_2)
\!\ddd\! O(w_{2n},\bar{w}_{2n})\right\rangle_{\Sigma_n}}
\!\!-\!\!n\log\left\langle{O}^{\dagger}(w_l,\bar{w}_1)O(w_2,\bar{w}_2)\right\rangle_{\Sigma_1}\!\!\Biggr] \ \ \ \ \label{replica}
\ea
where $(w_{2k+1},w_{2k+2})$ for $k=1,2,\ddd,n-1$ are $n-1$ replicas of $(w_1,w_2)$ in the $k$-th sheet of $\Sigma_n$. The first term in the second line is given by a $2n$ point correlation function on $\Sigma_n$. The second term is a two point function on $\Sigma_1=R^2$ and we normalized this such that
\ba
\la O^{\dagger}(w_1,\bar{w}_1)O(w_2,\bar{w}_2)\lb_{\Sigma_1}
\!=\!|w_{12}|^{-4\Delta_O}\!=\!(2\ep)^{-4\Delta_O}, \label{ntw}
\ea
which is equal to ${\mathcal N}^{-1}$.

Now we would like to focus on the second Renyi entropy $\Delta S^{(2)}_A$. Please refer to
\cite{He:2014mwa} for more details. Let us apply the conformal transformation:
\be
w/(w - L) = z^n , \label{cmap}
\ee
which maps $\Sigma_n$ into $\Sigma_1$. By employing (\ref{wco}) and (\ref{wcor}) with $n=2$, we find that  $(z_i,\bar{z}_i)$ are given
\ba
&& z_1 = -z_3 = \sqrt{(l - t - i \ep)/(l + L - t - i \ep) } , \no
&& z_2 = -z_4 = \sqrt{(l - t + i \ep)/(l + L - t + i \ep ) } . 
\ea
Moreover, the cross ratios are $z = z_{12}z_{34}/(z_{13}z_{24}),\ \bar{z} = \bar{z}_{12}\bar{z}_{34}/(\bar{z}_{13}\bar{z}_{24})$,
where $z_{i j} = z_i - z_j$.  Consider now the behavior of $(z , \bar{z})$  in the limit $\ep\to 0$. When $0<t<l$ or $t>L+l$, one can show that $(z,\bar{z})\to(0,0)$ as
\be
z\simeq\f{L^2\ep^2}{4(l-t)^2(L+l-t)^2},\ \  \ \ \bar{z}\simeq\f{L^2\ep^2}{4(l+t)^2(L+l+t)^2}.
\label{etim}
\ee
In the other case $l<t<L+l$, we find $(z,\bar{z})\to(1,0)$:
\be
z\simeq 1-\f{L^2\ep^2}{4(l-t)^2(L+l-t)^2},\ \  \bar{z}\simeq \f{L^2\ep^2}{4(l+t)^2(L+l+t)^2}.\label{lik}
\ee
 Though the limit $(z,\bar{z})\to(1,0)$ does not satisfy the complex conjugation, it inevitably arises due to the analytical continuation of $t$.

Owing to the conformal symmetry, the four point function on $\Sigma_1$ can be written as
\ba
&& \la O^{\dagger}(z_1,\bar{z}_1)O(z_2,\bar{z}_2) O^{\dagger}(z_3,\bar{z}_3) O(z_4,\bar{z}_4)\lb_{\Sigma_1} \no
&& =|z_{13}z_{24}|^{-4\Delta_O}\cdot G(z,\bar{z}). \label{wgf}
\ea
Applying the conformal map (\ref{cmap}), we obtain the four point function on $\Sigma_2$:
\ba
&& \la O^{\dagger}(w_1,\bar{w}_1)O(w_2,\bar{w}_2)O^{\dagger}(w_3,\bar{w}_3)O(w_4,\bar{w}_4)\lb_{\Sigma_n}\no
&& \!=\!\prod_{i=1}^{4}\left|dw_i/dz_i\right|^{-2 \Delta_O }\!
\la O^{\dagger}(z_1,\bar{z}_1)O(z_2,\bar{z}_2)O^{\dagger}(z_3,\bar{z}_3)O(z_4,\bar{z}_4)\lb_{\Sigma_1} \no
&& = (4L)^{-8\Delta_O} \left|(z_1^2 - 1) (z_2^2 - 1)/(z_1z_2) \right|^{8\Delta_O} \cdot G(z,\bar{z}).
\ea
Using this and (\ref{ntw}), the relevant ratio that we need to compute $\Delta S^{(2)}_A$  is expressed as 
\ba
&&  \f{\la O^{\dagger}(w_1,\bar{w}_1)O(w_2,\bar{w}_2)O^{\dagger}(w_3,\bar{w}_3)O(w_4,\bar{w}_4)
\lb_{\Sigma_n}}{\left(\la O^{\dagger}(w_1,\bar{w}_1)O(w_2,\bar{w}_2)\lb_{\Sigma_1}\right)^2} \no
&& =|z|^{4\Delta_O}|1-z|^{4\Delta_O} \cdot G(z,\bar{z}) . \label{fgf}
\ea

For example, consider a $c=1$ CFT defined by a massless free scalar $\phi$. Let us choose operators
\be
O_1 = e^{\f{i}{2} \phi},\ \ \ O_2=(e^{\f{i}{2} \phi} + e^{-\f{i}{2} \phi})/\s{2}, \label{opew}
\ee
which have a common conformal dimension $\Delta_1=\Delta_2 = \f{1}{8}$.
The functions $G(z , \bar{z})$ for $O_1$ and $O_2$ can be computed as follows
\ba
&& G_1(z , \bar{z}) = 1/\sqrt{ \left| z \right| \left| 1 - z \right| } ,\no
&& G_2(z , \bar{z})= (|z| + 1 + |1 - z|)\cdot G_1(z , \bar{z})/2 ,
\ea
respectively. It is obvious that for the operator $O_1$ the Renyi entropy vanishes at all times $\Delta S^{(2)}_A=0$. For $O_2$, taking into account the two periods (\ref{etim}) and (\ref{lik})
we get
\be
\Delta S^{(2)}_A =
\left\{
\begin{array}{l}
 \  0  \ \ \ \ \ \ \ \ \left(0 <  t < l, \ \mbox{or}\  \  t> l + L   \right) , \\
 \ \log 2 \ \ \ \ \ \ \ \ \left(l < t  < l + L  \right) .
\end{array}
\right  . \label{spp}
\ee

Now let us give a heuristic interpretations of these results in the $c=1$ CFT. First of all it is clear that we will always get the trivial result $\Delta S^{(n)}_A$ for the time
$0<t<l$ and $t>l+L$ \cite{He:2014mwa} (we will review this proof in section 4.1). This is because the local operator creates an entangled pair(s) at $x=-l$ and each half of the pairs propagate in the opposite direction at the speed of light. However for the time  $0<t<l$ and $t>l+L$, all of them are inside the subsystem B. Therefore they do not contribute to the quantum entanglement between $A$ and $B$.

On the other hand, during the time $l<t<t+L$, $\Delta S^{(n)}_A$ can be non-trivial because a half of of the entangled pairs is situated in $A$ and the other half is in $B$. The ones in $A$ (and $B$) consists of the right-moving (and left-moving) mode, respectively.

Now we obtain the trivial result for $O_1$ even during the time $l<t<t+L$. This is because the excited state $e^{\f{i}{2}\phi}|0\lb$ can be regarded as a direct product state
\be
e^{\f{i}{2}\phi_L}|0\lb_L\otimes e^{\f{i}{2}\phi_R}|0\lb_R,
\ee
in the left-moving (L: chiral) and right-moving (R: anti-chiral) sector \cite{NNT}. This is, of course, not an entangled state. On the contrary, $O_2$ creates a maximally entangled state (or equally EPR  state):
\be
\f{1}{\s{2}}\left(e^{\f{i}{2}\phi_L}|0\lb_L\otimes e^{\f{i}{2}\phi_R}|0\lb_R+
e^{-\f{i}{2}\phi_L}|0\lb_L\otimes e^{-\f{i}{2}\phi_R}|0\lb_R\right),
\ee
 which produces the Renyi entanglement entropy $\log 2$ for any $n$. This explains the behavior (\ref{spp}).

We will continue the analysis of $\Delta S^{(n)}_A$  in general cases including large central charge CFTs in section 4.

\subsection{Analysis of Energy Density}

Before we go on, we would like to briefly comment on the behavior of energy density $T_{tt}$.
For simplicity, consider a 2d CFT on the 2d Euclidean flat space $R^2=\Sigma_1$ and employ the coordinate
 $(w,\bar{w})=(x+i\tau,x-i\tau)$ defined in the previous subsection. We create an excited state by inserting a primary operator $O$ at $x=-l<0$ so that the time evolution of the density matrix looks like
 (\ref{dmtd}).

In this setup, the energy density of this excited state at $(z,\bar{z})=(x,x)$  after the time $t$ can be found as
\be
\la T_{tt}\lb =\f{\la O^\dagger(w_2,\bar{w}_2)T_{tt}(x,x)O(w_1,\bar{w}_1)\lb}{\la O^\dagger(w_2,\bar{w}_2)O(w_1,\bar{w}_1)\lb}
=\Delta_O \ep^2\left[\f{1}{\left((x+l-t)^2+\ep^2\right)^2}
+\f{1}{\left((x+l+t)^2+\ep^2\right)^2}\right], \label{energyd}
\ee
where $(w_i,\bar{w}_i)$ were already defined in (\ref{wco}) and (\ref{wcor}).

Note that this formula is true for any primary operator $O$ as the three point function including an energy stress tensor is universal in 2d CFTs. This result (\ref{energyd}) manifestly shows that the energy density is localized at the two points $x=-l\pm t$, which is simply explained by a relativistic propagation of energy.
Since we are interested in the dynamics in the subsystem $A$ after the excitation $t>0$, only the point $x=-l+t$ is relevant.

Thus the excitation is included in the subsystem $A$ when $l-L/2<t<l+L/2$. During this time,
the total energy increase $\Delta E_A$ can be estimated as follows:
\be
\Delta E_A=\int_{x\in A} dx T_{tt}(x)\sim \f{\Delta_O}{\ep}, \label{eaer} \\
\ee
up to an overall factor. This gets divergent in the point-like limit $\ep\to 0$.

One may think it will be useful if we can find any direct relation between the excitation energy and the entanglement entropy growth. In particular cases, such a relation is known and is called the first law-like relation. Here the first law means a linear relation between the change of entanglement entropy: $\Delta S^{(1)}_A$ and that of energy $\Delta E_A=\int_{x\in A} dx^d T_{tt}(x)$ or energy density $\Delta T_{tt}$ in the subsystem $A$. The first law-like relation was first found in \cite{BNTU} by using the holographic entanglement entropy \cite{RT} for any choice of the subsystem $A$, assuming the translational invariance. This relation is expressed as follows
\be
\Delta E_A\simeq T_{A}\cdot \Delta S^{(1)}_A,
\ee
where the `effective temperature' $T_{A}$ scales like $T_A=\f{C_A}{L}$ using the linear size $L$ of
the subsystem $A$. The coefficient $C_A$ depends on the shape of $A$. We can apply this formula only for small
excitations such that $L^{d+1}T_{tt}\ll 1$. Refer also to \cite{Firstlaw} for more holographic calculations.

 Remarkably this was extended to general inhomogeneous cases in \cite{BCHM,WKPW} by employing the analysis of conformal transformation \cite{CHM}. When the subsystem $A$ is given by a round ball $|x|\leq L/2$ with the radius $L/2$, the following relation
\be
\Delta S^{(1)}_A\simeq \int_{|x|\leq L/2} dx^d \left(\f{L^2/4-|x|^2}{L}\right) T_{tt}(x),
\label{fst}
\ee
was shown for small excitations in any $d+1$ dimensional CFT even without assuming holography.

Therefore one might think that we may compute $\Delta S^{(1)}_A$ from the energy increase by using this first law-like relation. However, this is not the case as we explain below.
Indeed, when $l-L/2<t<l+L/2$, the right-hand side of (\ref{fst}) can be estimated as follows:
\ba
&& \int_{|x|\leq L/2} dx \left(\f{L^2/4-|x|^2}{L}\right) T_{tt}(x)\sim \f{\Delta_O (L^2/4-(t-l)^2)}{L\ep}.
\ea
up to an overall factor.

If we naively apply the first law (\ref{fst}), we will be in trouble because $\Delta S^{(1)}_A$ gets divergent in the $\ep\to 0$ limit. This contradicts the fact that in 2d rational CFTs \cite{He:2014mwa} $\Delta S^{(1)}_A$ should be equal to the log of the quantum dimension which is finite. Indeed we can easily see why the first law (\ref{fst}) breaks down because the energy $\Delta E_A$ gets divergent as in (\ref{eaer}) and thus should not be regarded as a small excitation \footnote{See \cite{Giusto:2014aba} for a similar conclusion in time independent setup.}. Nevertheless we would like to mention that if we keep $\ep$ to be large, then we can apply the first law because the energy density gets very small, though we are not interested in such a situation in this paper.

\section{Excited States in Large N Free Yang-Mills}

Here we would like to compute the $n$-th Renyi entanglement entropy in free $U(N)$ gauge theory in four dimension. We are interested in excited states which are obtained by acting the following gauge invariant local operators:
\begin{equation}
\Tr \mathcal{Z}^J =\Tr (\phi_1+i\phi_2)^J,  \label{tropl}
\end{equation}
on the vacuum. Here $\phi_{1}$ and $\phi_2$ represent two real massless scalar fields in $R^{1,3}$, which belongs to the adjoint representation. In this work we assume that $J\sim O(1)$ but it will be an interesting future problem to repeat our analysis for operators with large $J$ (of order of some powers of $N$).

\subsection{Replica Method Calculations}

It is rather complicated to directly compute the $n$-th Renyi entanglement entropy for this state
by using the replica method formula (\ref{gre}). However, if we take the $\epsilon \rightarrow 0$ and the large $N$ limit ($N \rightarrow \infty$), the computation gets drastically simplified.

 In the large $N$ limit, the leading contributions come from only disconnected planar diagrams. Then the two point function on $\Sigma_n$ is approximated by\footnote{Here we ignore the difference between $SU(N)$ and $U(N)$ gauge theory as we take large $N$ limit.}
\begin{equation}
\left\langle Tr \mathcal{Z}^J(w, \bar{w},{\bf x}) Tr \bar{\mathcal{Z}}^J(w', \bar{w}', {\bf x'})\right \rangle =J N^{J} (G_{n})^J+(1/N\ \mbox{corrections}),
\end{equation}
where the Green function $G_n$ is defined by (\ref{gfnwq}) in section 2.

In the large $N$ limit, the $2n$ point function in the replica formula (\ref{gre}) is at most $\mathcal{O}(N^{J n})$.
 Various disconnected diagrams of planar diagrams can contribute to the $2n$ point function of $Tr \mathcal{Z}^J$. However, we can show that only two disconnected diagrams, which are depicted in Fig.3 (for n=4 example), can contribute to  $\Delta S^{(n)}_A$ in the limit $\ep\to 0$ as pointed out in \cite{Nozaki:2014uaa}. 
\begin{figure}[h]
 \begin{minipage}{0.5\hsize}
  \begin{center}
   \includegraphics[width=73mm]{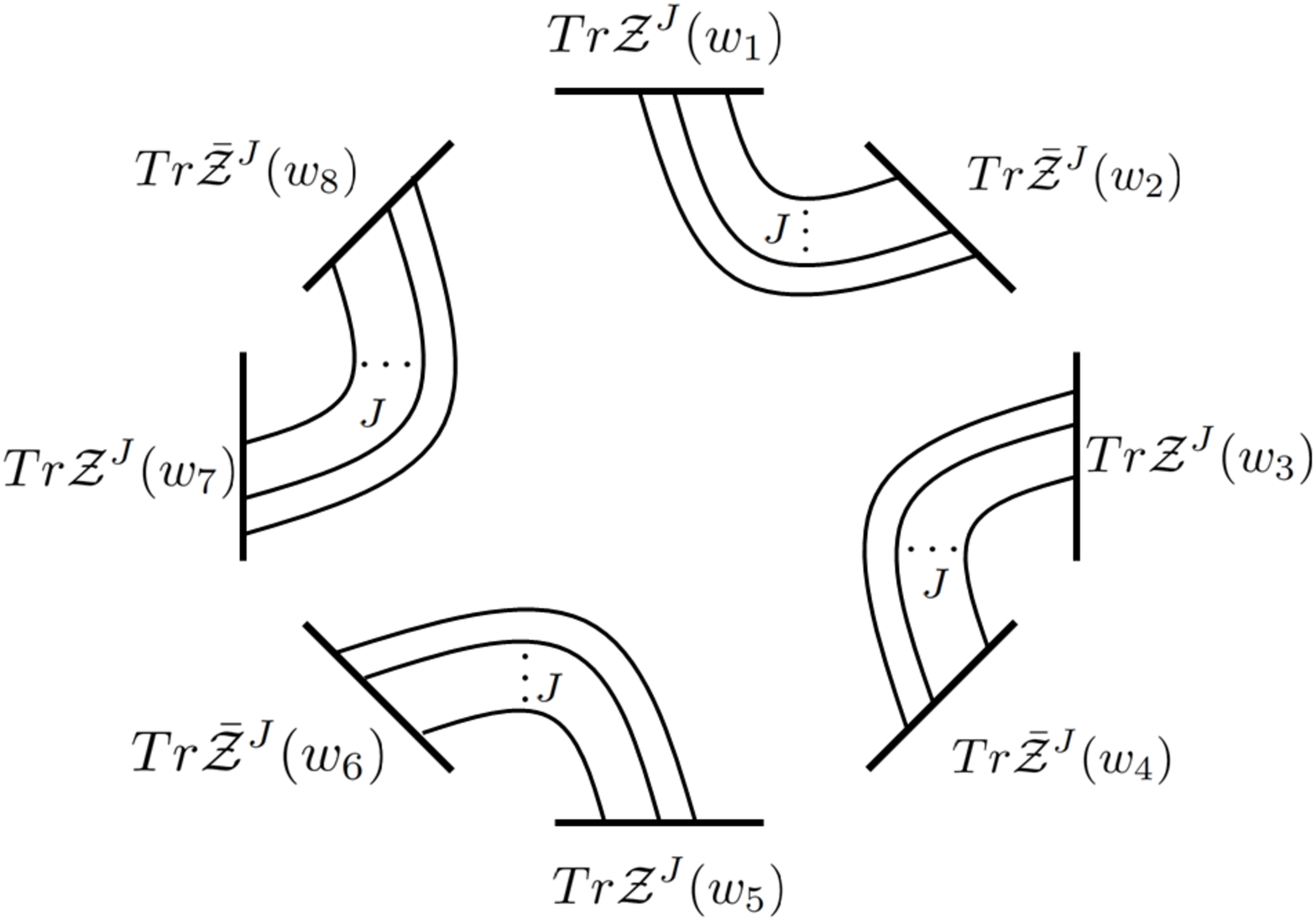}
  \end{center}
 \end{minipage}
  \begin{minipage}{0.30\hsize}
  \begin{center}
   \includegraphics[width=73mm]{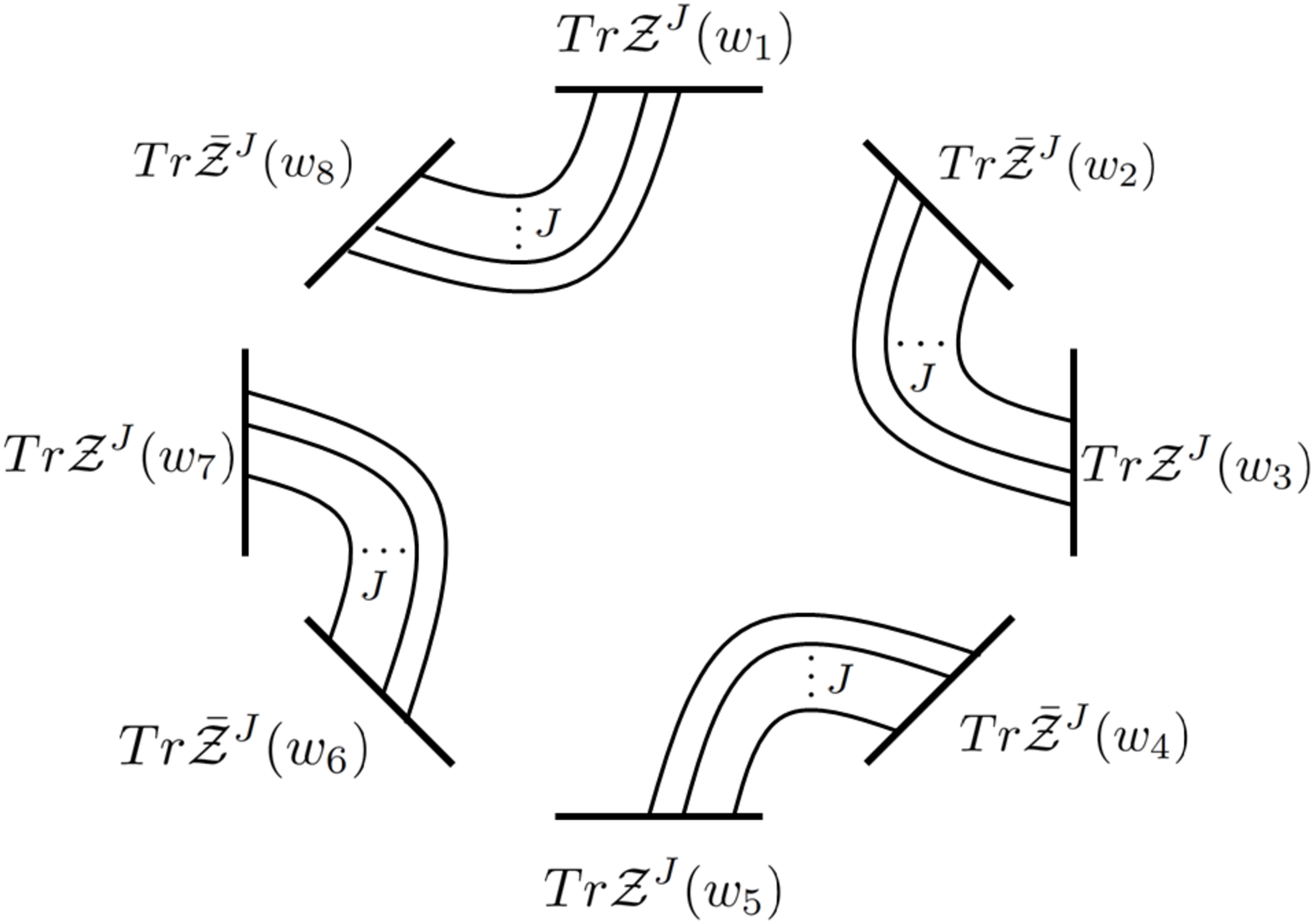}
  \end{center}
 \end{minipage}
 \caption{The two dominant diagrams for the $4$th Renyi entanglement entropy $\Delta S^{(4)}_A$  in the large $N$ and $\epsilon \rightarrow 0$ limit. }\label{ph}
\end{figure}

One of them corresponds to the product of the propagators between two points on same sheet
(remember Fig.1). Another one is the product of the propagators between $w_{2k}$ and $w_{2k+1}$, where
$k$ runs from $1$ to $n$.

In this way, the leading term of $\Delta S^{(n)}_A$ is computed as
\begin{eqnarray}
\Delta S^{(n)}_A=
\left\{ \begin{array}{ll}
0 & t < l, \\
\frac{\log{\left[\frac{ J^n N^{n J}\left(\frac{l+t}{32 \pi ^2 t \epsilon ^2}\right)^{n J}+ J^n N^{n J}\left(\frac{-l+t}{32 \pi ^2 t \epsilon ^2}\right)^{n J}}{J^n N^{n J}\left(\frac{1}{16 \pi ^2 \epsilon ^2}\right)^{n J}}\right]} }{1-n}
=-\frac{\log{\left[2^{-J n} \left(\left(1-\frac{l}{t}\right)^{J n}+\left(\frac{l+t}{t}\right)^{J n}\right)\right]}}{-1+n}& t \ge l .\\
\end{array} \right.  \label{ldwqw}
\end{eqnarray}

\begin{figure}[h]
  \centering
  \includegraphics[width=8cm]{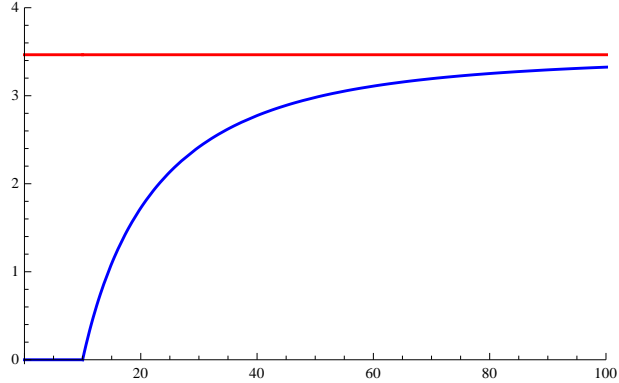}
  \caption{This is the plot of $\Delta S^{(2)}_A$ (blue) as function of $t$ in the $\epsilon \rightarrow 0$ limit.
Here we chose $l=10$, $n=2$ and $J=3$. The red line corresponds to the late time value $5\log{2}$.}\label{l10n2j3}
\end{figure}

Moreover, if we take the late time limit $t \rightarrow \infty$, the final value of Renyi entropy is given by
\begin{equation}
\Delta S^{(n)}_A=\frac{(J n-1) \log{2}}{n-1}. \label{free}
\end{equation}
The time evolution of $\Delta S^{(2)}_A$ for $J=3$ is plotted in Fig.\ref{l10n2j3}. We immediately notice that  $\Delta S^{(n)}$ diverges in the von-Neumann entropy limit $n=1$. As we explain later by using a different method, this problem occurs because we ignore
the higher order terms in the large $N$ expansion which becomes important only in $n=1$ limit. We can find from this observation that we can trust the leading order result (\ref{ldwqw}) iff $n>1$.

\subsection{Left-Right Decomposition and Entangled Pair Interpretation}

There is another method to compute the late time limit of $\Delta S^{(n)}_A$ by using the structures of excited states \cite{Nozaki:2014hna,Nozaki:2014uaa}.
We can regard an excited state produced by a local operator in a CFT as a combination of creation operators acting on the left-moving vacuum $|0\lb_L$ and the right-moving one $|0\lb_R$ in the product Hilbert space $\HH_L\otimes \HH_R$. Thus, when we insert the local operator, this creates an entangled pair and each half of the pair will propagate in the left or right direction. At late time, one of them will be in the subsystem $A$ and the other in the subsystem $B$. Thus the excited state will have a non-trivial quantum entanglement at late time. Therefore, we can
compute the Renyi entropy $\Delta S^{(n)}_A$ directly from the reduced density matrix obtained by tracing out the left-moving (or right-moving) part.

In this section we apply this analysis to our trace operators (\ref{tropl})
built out of matrices in four dimensional CFTs with $U(N)$ gauge group. By decomposing
the $U(N)$ scalar field $\phi$ into the left and right-moving mode as in (\ref{lrdecf}), we can express our excited states as follows:
\be
\left|\Psi_{\Tr[\phi^J]}\right\lb\equiv \Tr\left[(\phi_L+\phi_R)^J\right]\left|0_L\right\lb\left|0_R\right\lb,
\ee
where we take $\phi_{L/R}$ to be $N\times N$ matrices with the Wick contraction rule
\be
\la\left(\phi_A\right)^\alpha_\beta\left(\phi_B\right)^\mu_\nu\lb
=\delta_{AB}\,\delta^\alpha_\nu\delta^\mu_\beta,\qquad A,B\in\{L,R\}\label{UN}
\ee
The corresponding density matrix in $\HH_{L}\otimes \HH_R$ space is
\be
\rho_{L\cup R}=\ket{\Psi_{\Tr[\phi^J]}}\bra{\Psi_{\Tr[\phi^J]}}
\ee
We can then trace over $\HH_L$ and, using the replica method, evaluate the Renyi entropy of $\rho_R=\Tr_L\left(\rho_{L\cup R}\right)$ as
\be
\Delta S^{(n)}_R=\frac{1}{1-n}\log \Tr \rho^n_R
\ee
As a warm up, consider the $J=1$ case in which the excited state is simply
\be
\ket{\Psi_{\Tr[\phi]}}\propto\left(\ket{\Tr(\phi_L)}\ket{0_R}+\ket{0_L}\ket{\Tr(\phi_R)}\right).
\ee
The orthonormal basis in $\HH_L$  (also $\HH_R$) has two normalized states, the vacuum and a state obtained by acting with a single trace
\be
\left\{\ket{0_{L}},\f{1}{\s{N}}\ket{\Tr(\phi_{L})}\right\}
\ee
where the normalization of the single trace state can be deduced from \eqref{UN}. Tracing over $\HH_L$ gives the reduced density matrix in $\HH_R$
\be
\rho_R=\f{1}{2}\ket{0}\bra{0}+\f{1}{2N}\ket{\Tr(\phi_{R})}\bra{\Tr(\phi_{R})}
\ee
which after normalization to $\Tr (\rho)$=1 becomes $\rho_R=\text{diag}\left(\frac{1}{2},\frac{1}{2}\right)$. Finally, the $n$-th Renyi entanglement entropy for the state excited by $\Tr(\phi)$ is given by
\be
\Delta S^{(n)}_R=\frac{1}{1-n}\log\left(2^{1-n}\right)=\log(2)
\ee
It is easy to see that $\ket{\Psi_{\Tr[\phi]}}$ in $\HH_L\otimes \HH_R$ is equivalent to the EPR state, therefore all the Renyi entropies are equal to $\log 2$, which is the entropy of the maximally entangled state. Note that this result is same as the $N=1$ case reviewed in section 2.1. This is simply because we can treat $\Tr[\phi]$ as a single real scalar field by diagonalizing the Hamiltonian of the system.

The situation becomes more interesting once we consider traces of products of matrices. For $J=2$ the excited state is given by
\be
\ket{\Psi_{\Tr[\phi^2]}}\propto\ket{\Tr\phi^2_L}\ket{0_R}+\ket{0_L}\ket{\Tr\phi^2_R}+2\sum^{N}_{\alpha,\beta=1}\ket{\left(\phi_L\right)^\alpha_\beta}\ket{\left(\phi_R\right)^\beta_\alpha}
\ee
where in the third term we explicitly wrote the indices in $\Tr(\phi_L\phi_R)\ket{0_L}\ket{0_R}$.\\
The Hilbert space $\HH_L$ consists of the vacuum, the state obtained by action of $\Tr(\phi^2_L)$ and $N^2$ states $\ket{\left(\phi_L\right)^{\alpha}_\beta}$, all with appropriate normalizations. Tracing out the $\HH_L$ and normalizing gives the reduced density matrix in $\HH_R$
\be
\rho_R=\text{diag}\left(\frac{1}{4},\frac{1}{4},\frac{\vec{1}}{2N^2}\right)
\ee
where $\vec{1}$ is the $N^2$ dimensional unit vector. Using the replica trick we again obtain the $n$-th Renyi entropies
\bea
\Delta S^{(n)}_R=\frac{1}{1-n}\log\left(2^{1-2n}+\frac{1}{2^nN^{2(n-1)}}\right). \label{subldf}
\eea

Notice now the striking difference between Renyi entropies with $n\ge 2$ and the von-Neumann entropy for $n=1$. For $n\ge 2$ and large $N$, we can simply neglect the $1/N$ corrections and the leading result is
\be
\Delta S^{(n\ge2)}_R\simeq \frac{2n-1}{n-1}\cdot\log 2.
\ee
This perfectly matches with \eqref{free} obtained in the large $N$ limit using free field Wick contractions.

However, in the limit $n=1$, the von-Neumann entropy is equal to
\be
\Delta S^{(1)}_R=\log\left(2\sqrt{2}N\right),
\ee
where we took into account the sub-leading term. The reason why the subleading term in the large $N$ expansion becomes important is because precisely when $n\to 1$ it overwhelms the originally leading term. In other words, taking the large $N$ limit first assuming $n>1$ and after that evaluating the entanglement entropy for excited states leads to an incorrect result. This way, we find an interesting breakdown of the naive large $N$ limit\footnote{Note that this breakdown occurs because we neglect connected planar diagrams. A similar breakdown of perturbation theory has been observed in a wide range of examples of calculating von-Neumann entanglement entropy under relevant operator perturbations. This was pointed out in a specific example in \cite{SiTa} (see eq.(2.43) and (2.44) in that paper), which typically leads to logarithmic contributions. For recent progresses of analysis of entanglement entropy in the presence of operator perturbations refer to e.g.\cite{pertab}.} that occurs for the von-Neumann entropy and enhances it by a factor $O(\log N)$.

The analysis becomes more cumbersome for higher $J\sim O(1)$ (see App.\ref{HigherJ}) but assuming $n\ge 2$ we can verify that at large $N$
\be
\Tr \rho^n_R=2^{1-Jn}+\frac{(J-1)J^n}{2^{Jn}N^{2(n-1)}}+O(1/N^{4(n-1)})
\ee
and the Renyi entropies\footnote{Even though we assumed $J\sim O(1)$, it is interesting to note that the second term becomes important once $J\sim N^{2\frac{n-1}{n+1}}$.}
\be
\Delta S^{(n)}_R\sim \frac{J\,n-1}{n-1}\log 2
\ee
perfectly match the leading order of the replica calculation \eqref{free}. It is also interesting to point that the min-entropy $(n\to\infty)$ shows a universal behavior
\be
\Delta S^{(\infty)}_R\sim J\log 2
\ee

For the von-Neumann entropy, we cannot neglect the $1/N$ corrections and we need to take into account other terms which can contribute as $n\to 1$. Nevertherless, we can again estimate it as
\be
\Delta S^{(1)}_R\sim J\log N,  \label{vonjn}
\ee
up to a certain $O(1)$ constant.

It is intriguing that we observe the $\log N$ entropy only for the von-Neumann entropy. The scalar field is a $N\times N$ matrix valued and its almost maximally entangled state can produce the $\log N$ entropy. Furthermore, since there are $J$ scalar operators in Tr$[\phi^J]$, then we can easily understand (\ref{vonjn}). However, a lesson we learned here is that this naive argument cannot be applied to the Renyi entropies. If we roughly regard $n$ as the inverse temperature, we might say that the result of von-Neumann entropy $n=1$ looks `deconfined', while
those for Renyi entropies $n=2,3,\ddd$ looks `confined' as the former corresponds to higher temperature than the latter. We hope to understand this ``transition"-like behaviour better in our future work.

\section{Two dimensional CFT at Large $c$}

Now we turn to two dimensional CFTs (2d CFTs) having in mind the AdS$_3/$CFT$_2$ correspondence. We will follow the formulation explained in section 2.2. In order to have a possibility of a classical gravity dual, we are interested in taking the large $c$ limit, where $c$ is the central charge of a CFT.

\subsection{General Arguments}

The function $G(z,\bar{z})$ defined in (\ref{wgf}) is the essential part of the four point function. Owing to the holomorphic/anti-holomorphic factorization, this function
can be written as a sum of conformal blocks
\cite{BPZ}:
\be
G(z,\bar{z})=\sum_{b}(C^{b}_{OO^\dagger})^2F_O(b|z)\bar{F}_O(b|\bar{z}). \label{cbl}
\ee
The index $b$ corresponds to each $\phi_b$ of all Virasoro primary fields and note that they are decomposed
into the left-moving (chiral) and right-moving (anti-chiral) primary fields. Thus we can
write $b$ as $(b_L,b_R)$, corresponding to $\phi_b(z,\bar{z})=\phi_{b_L}(z)\phi_{b_R}(\bar{z})$.
We assume the summation over $b$ is a discrete sum, as is true in typical AdS/CFT examples.

How to take this summation of $b=(b_L,b_R)$ depends on the spectrum of 2d CFT we consider.
One simple class of CFTs, such as the rational CFTs, is defined by the diagonal sum $b_L=b_R(\equiv a)$, where the total Hilbert space is given by
\be
{\cal H}=\oplus_{a}\left({\cal H}^{(a)}_{L}\otimes {\cal H}^{(a)}_{R}\right). \label{diagcft}
\ee
However, in general, this is not the case because there are generators other than the Virasoro
operators. For example, if consider the current algebra or equally WZW model, (\ref{diagcft}) is correct only if we regard the index $a$ as that for all primary fields of current algebra instead of Virasoro algebra.

The conformal block $F_O(b|z)$ is normalized such that in the $z\to 0$ limit we have
\be
F_O(b|z)\to z^{\Delta_b-2\Delta_O}, \label{bgsq}
\ee
where $\Delta_b$ is the conformal dimension of the primary operator labeled by $b$ (called $\phi_b$). The factor $C^{b}_{OO^\dagger}$ is the coefficient of normalized three point function $\la OO^\dagger\phi_b\lb$.

First we consider the time $0<t<l$ or $t>l+L$, where localized excitations which are generated at $x=-l$ and propagate in the opposite directions at the speed of light, are outside of the subsystem $A$. This means that the entangled pairs are both in the region $B$. Therefore we expect that there is no entanglement generation between $A$ and $B$ as we already mentioned in section 2.2. Indeed, we can obtain from (\ref{etim}) that in the $\ep\to 0$ limit we have $(z,\bar{z})\to (0,0)$ and $G(z,\bar{z})\simeq |z|^{-4\Delta_O}$, as the dominant contribution arises when $\phi_b$ coincides with the identity operator $\phi_0=I$.
Thus (\ref{fgf}) leads to $\Delta S^{(2)}_A=0$, as expected.

The more interesting time $l<t<l+L$, during when a half of the entangled pairs is in the region $A$ and the others in $B$, requires some knowledge of the behavior of the conformal block $F_O(b|z)$ in the limit $z\to 1$ as we will discuss below.

\subsection{Rational 2d CFTs}

Before we start the large $c$ analysis, we would like to briefly review the results in
rational CFTs obtained in \cite{He:2014mwa}. This is useful because they provide an instructive example where we can compute $\Delta S^{(n)}_A$ both analytically and exactly. In rational CFTs the summation over $b$ is a finite sum and we can choose the Hilbert space structure given by
(\ref{diagcft}).

In rational CFTs, we can employ the properties of the fusion transformation, which exchanges $z_2$ with $z_4$ (or equally $z$ with $1-z$) given by
\be
F_O(b|1-z)=\sum_{d}F_{bd}[O]\cdot F_O(d|z), \label{Ftr}
\ee
where the coefficients $F_{bd}[O]$ are constants, which are called fusion matrices \cite{MSP}.
Thus by using (\ref{bgsq}), in the limit $(z,\bar{z})\to (1,0)$, the conformal block is reduced to the contribution from the vacuum sector:
\be
G(z,\bar{z})\simeq F_{00}[O]\cdot (1-z)^{-2\Delta_O}\bar{z}^{-2\Delta_O},
\ee
where we employed the fact that $C^{0}_{OO}=1$.

Therefore we find the following expression from (\ref{fgf}):
\be
\Delta S^{(2)}_A=-\log F_{00}[O]=\log d_O,
\ee
where $d_O=1/F_{00}[O]$ is called the quantum dimension \cite{MSP} and is related to the S-matrix of the modular transformation by
\be
d_O=S_{0O}/S_{00}. \label{qdm}
\ee

\subsection{Large $c$ limit}

Now we move on to the large $c$ limit of 2d CFTs. We are interested in the time period $l<t<L+l$, where we expect non-trivial results, corresponding to the limit $(z,\bar{z})\to (1,0)$ as explained in (\ref{lik}).
We will keep only the leading order of $\f{\Delta_O}{c}(\ll 1)$ expansion. We will discuss sub-leading corrections in the subsection 4.5 later.

Since we are motivated by the AdS/CFT, we are interested in those CFTs with gravity duals. Therefore we would like to assume the existence of the gap in the spectrum such that the density of states $d(\Delta)$ behaves like $d(\Delta)\sim O(1)$ for $\Delta<O(c)$. This corresponds to the threshold where AdS black holes appear. Moreover, in the summation of conformal blocks (\ref{cbl}) we can ignore the contributions from intermediate states with large conformal dimension $\Delta_b\sim O(c)$, as their conformal blocks are exponentially suppressed in the large $c$ limit \cite{Fateev:2011qa,Hartman}.

These arguments are parallel with the paper \cite{Hartman}, where the ground state entanglement entropy in large $c$ limit was analyzed. However, note that in that paper, the large $c$ limit was taken with $\f{\Delta_O}{c}$ kept finite because the correlation functions of twist operators were computed. In our case, the operator $O$ expresses the excitation above the vacuum and we do not need any twist operators as we employed the conformal map to describe the replicated Riemann surface $\Sigma_2$.

In our large $c$ limit $\f{\Delta_O}{c},\f{\Delta_b}{c}\ll 1$, we have the following simple and universal expression of the vacuum conformal block \cite{Zam,Fateev:2011qa}:
\be
F_O(b|z)\simeq z^{\Delta_b-2\Delta_O}\cdot{}_2F_1(\Delta_b,\Delta_b,2\Delta_b,z), \label{lcd}
\ee
where ${}_2F_1(a,b,c,z)$ is the hypergeometric function. This shows that for any $\Delta_b\ll c$ the conformal block $F_O(b|z)$ can only possess at most a logarithmic singularity $\sim \log (1-z)$ in the limit
$z\to 1$.

However, to perform a complete analysis of large $c$ limit, we need to know the structure of Hilbert space to deal with the summation over $b$ in (\ref{cbl}). Since this is in general too complicated, we would like to approximate it only by the contribution from the vacuum conformal block:
\be
G(z,\bar{z})\simeq F_O(0|z)\bar{F}_O(0|\bar{z}). \label{apgf}
\ee
This argument is clearly justified for the diagonal Hilbert space structure (\ref{diagcft}) in the limit $(z,\bar{z})\to (1,0)$, though in general AdS/CFT examples we cannot expect this. This is because in this limit we have $\bar{F}_O(0|\bar{z})\simeq \bar{z}^{\Delta_b-2\Delta_O}$, which gives the dominant contribution if $\Delta_b=0$.

Moreover, we would like to argue that this approximation can be applied to examples with gravity duals.\footnote{Note that we cannot apply this argument to theory without gravity duals. For example, consider the free CFT defined by $c$ free massless scalar fields $\phi_1,\phi_2,\ddd,\phi_c$. Let us focus on the operator $O=e^{ip\phi_1}$ for any $p$. This clearly has the vanishing entropy $\Delta S^{(2)}_A=0$ and this does not agree with our later results e.g. (\ref{LargeC2}). In such free CFTs there is no gap in the spectrum and we need to sum up so many conformal blocks.} This is because we can set $b_R=0$ in the limit $\bar{z}\to 0$ and because the summation over $b_L$ just multiplies some factor which is not exponentially large with respect to the central charge.

In the language of AdS/CFT correspondence, this approximation (\ref{apgf}) corresponds to ignoring backreactions of local excitations to supergravity fields in the bulk AdS. We will confirm this in the next section explicitly.

Under this approximation (\ref{apgf}), we can simply evaluate $G(z,\bar{z})$ as follows:
\be
\lim_{(z,\bar{z})\to (1,0)}G(z,\bar{z})\simeq \bar{z}^{-2\Delta_O}.
\ee

By using (\ref{fgf}) and (\ref{lik}),  this leads to the following second Renyi entropy when $l<t<L+l$:
\ba
\Delta S^{(2)}_A&=&-\log (1-z)^{2\Delta_O},\no
&=& 4\Delta_O\cdot \log\left[\f{2(t-l)(L+l-t)}{L\ep}\right].
\ea

In the limit $l\ll t\ll L$ we obtain
\be
\Delta S^{(2)}_A\simeq 4\Delta_O\cdot \log\f{2t}{\ep}=2\hat{\Delta}_O\cdot \log\f{2t}{\ep}.\label{LargeC2}
\ee
 Remember that $\Delta_O$ is the chiral conformal dimension and it is related to the usual conformal dimension $\hat{\Delta}_O$ (common in the definition in higher dimensions) via $\hat{\Delta}_O=2\Delta_O$.

However, we have to be careful with the large $c$ approximation (\ref{lcd}) we employed. Note that the bootstrap equation
\be
G(z,\bar{z})=G(1-z,1-\bar{z}), \label{bsrl}
\ee
, which should be obeyed by any 2d CFTs, requires in the $z\to 1$ limit $F_O(b|z)$ should generally behave like
\be
F_O(b|z)\simeq \f{1}{D^b_O}\cdot (1-z)^{-2\Delta_O}+\ddd,  \ \ \ \ (z\to 1),  \label{stfg}
\ee
where $D^b_O$ is a certain constant. The relation (\ref{bsrl}) requires
\be
\sum_{b_L,b_R}\f{(C^{b}_{OO^\dagger})^2}{D^{b_L}_O\cdot D^{b_R}_O}=1. \label{bfdgh}
\ee

These are indeed true in the fusion relation (\ref{Ftr}) in rational CFTs and in this case, $D^0_{O}$ coincides with the quantum dimension $d_O$. Since this term is missing in the large $c$ approximation (\ref{lcd}) and also in perturbative corrections which will be discussed in subsection \ref{dcc}, we can speculate that it should come from non-perturbative corrections like $\f{1}{D^b_O}\sim \exp(-c^{a})$ ($a$ is a certain positive constant) and thus
 $D^b_O$ are expected to be very large $D^b_O\to \infty$ in the limit $c\to \infty$.\footnote{
If we focus on the modes $b$ corresponding to the large conformal dimension $\Delta_b\gg c$, we can apply the Cardy's formula \cite{Cardyc} to estimate the number of primaries in the large $c$ limit. This is because we can ignore the difference between the numbers of primaries and those with their descendants in the large $c$ limit. This leads to the approximation of the density of states: $d(\Delta)\sim e^{2p\s{c\Delta}}$, where $p>0$ is a certain $O(1)$ constant. If we naively assume that the normalized three point functions are all $O(1)$, then the bootstrap constraint (\ref{bfdgh}) leads to the estimation that $D^{b}_O \gtrsim e^{p\s{c\Delta_b}}$ for $\Delta_b\gg c$. Therefore for the state $\phi_b$ with a small conformal dimension $\Delta_b\ll c$, we expect
$D^{b}_O \gtrsim e^{p'c}$ with another $O(1)$ constant $p'(>0)$. This is indeed exponentially large in the large $c$ limit.}  In other words, the large $c$ approximation (\ref{lcd})
will break down if $z$ is very close to $1$ such as $z\sim (D^b_O)^{-\f{1}{2\Delta_O}}$.

 If we take into account  (\ref{stfg}), we find the following behavior of $\Delta S^{(2)}_A$ (we took the limit $L\to \infty$):
\be
\Delta S^{(2)}_A =
\left\{
\begin{array}{l}
 \  4\Delta_O\cdot  \log\f{2(t-l)}{\ep}  \ \ \ \ \ \ \ \ \left(\ep\ll t-l\ll (D^0_O)^{\f{1}{4\Delta_O}}\ep \right) , \\
 \ \log D^0_O \ \ \ \ \ \ \ \ \left(t-l\gg  (D^0_O)^{\f{1}{4\Delta_O}}\ep \right) .
\end{array}
\right  . \label{sppqq}
\ee
In this way we found a logarithmic time evolution in an intermediate process. Note that this is typical only for the large $c$ limit. Indeed, if we look at the previous examples of rational CFTs or free CFTs, we find $D^0_O\sim O(1)$ and thus the logarithmic period disappears. In the next section, we will reproduce the same logarithmic behavior from the gravity side of AdS/CFT and will show that similar results can be obtained even in higher dimensions. It will be a very intriguing future problem to evaluate $D^0_O$ in some concrete models of 2d CFTs with gravity duals.

\subsection{Equivalence to Free Field Wick Contractions}

To generalize the previous analysis of the logarithmic behavior to $\Delta S^{(n)}_A$ for any $n$,
it is useful to note that our large $c$ approximation is equivalent to the computations of $2n$-point functions on $\Sigma_n$ using ``(free field) Wick contractions". For simplicity let us take the entangling region in this section to be the half line ($L\to\infty$ in previous sections), and use the map from $\Sigma_1$ to $\Sigma_n$
\be
w(z)=z^n.
\ee
We can find the two-point function of operators $O(w_i,\bar{w}_i)$ on $\Sigma_n$
\be
\langle O^\dagger(w_1,\bar{w}_1)O(w_2,\bar{w}_2)\rangle_{\Sigma_n}=n^{-4\Delta_O}|w_1w_2|^{\frac{2\Delta_O(1-n)}{n}}|w^{1/n}_1-w^{1/n}_2|^{-4\Delta_O}
\ee
Moreover, it can be shown (see \eqref{LTratio}) that the ratio of the two-point function on $\Sigma_n$ to the two point function on $\Sigma_1$ at late time is given by
\be
\frac{\langle O^\dagger(w_1,\bar{w}_1)O(w_2,\bar{w}_2)\rangle_{\Sigma_n}}{\langle O^\dagger(w_1,\bar{w}_1)O(w_2,\bar{w}_2)\rangle_{\Sigma_1}}\simeq \left(\frac{\epsilon}{n\sin\left(\frac{\pi}{n}\right)t}\right)^{2\Delta_O}.
\ee
Using the fact that leading divergent contribution to the correlation function comes from two types of Wick contractions as in the arguments in section 3.1, we get at late time
\be
\frac{\langle O^{\dagger}(w_1,\bar{w}_1)O(w_{2},\bar{w}_{2})...O^{\dagger}(w_{2n-1},\bar{w}_{2n-1})O(w_{2n},\bar{w}_{2n})\rangle_{\Sigma_n}}{\left(\langle O^\dagger(w_1,\bar{w}_1)O(w_2,\bar{w}_2)\rangle_{\Sigma_1}\right)^n}\simeq 2\left(\frac{\epsilon}{n\sin\left(\frac{\pi}{n}\right)t}\right)^{2n\Delta_O}.
\ee
That leads to the $n$-th Renyi entropy
\be
\Delta S^{(n)}_A\simeq\frac{1}{1-n}\log\left[2\left(\frac{\epsilon}
{n\sin\left(\frac{\pi}{n}\right)t}\right)^{2n\Delta_O}\right]
=\frac{2n\Delta_O}{n-1}\log\left(\frac{n\sin\left(\frac{\pi}{n}\right)t}{\epsilon}\right)-\frac{1}{n-1}\log 2.\label{DS3}
\ee
Setting $n=2$ we recover the logarithmic behavior at large $c$ \eqref{LargeC2}.

However, the same caution should be offered as previously. If we take $\ep\to 0$ strictly, our large $c$ approximation breaks down. In other words, if $t$ gets very large such that $t\gg (D^{0}_O)^{\f{1}{4\Delta_O}}$, the logarithmic growth will be terminated and $\Delta S^{(n)}_A$ approaches to some constant of order $\log D^{0}_O$ as in (\ref{sppqq}).

A related observation is that we cannot trust the results in the limit $n\to 1$ since not only $n\sin \pi/n\to0$ but the constant term blows-up when $n\to 1$. We believe that this problem occurs because our large $c$ approximation breaks down in the $n\to 1$ limit. A similar situation occurs in local operator excited states in free Yang-Mills as we studied
in section 3, where the large $N$ expansion breaks down in the limit $n\to 1$. We will comment more on this later by comparing with holographic results.

Before we proceed, it is useful to rewrite the ratio of the two-point functions on $\Sigma_n$ and $\Sigma_1$ in terms of coordinates $w=e^{\phi+i\tau/R}$. We will later compare it with the same ratio obtained holographically using geodesics in topological black hole. In $\{\tau,\phi\}$ coordinates we have
\bea
|w_1w_2|^{\frac{n-1}{n}}&=&e^{\phi_1+\phi_2-\frac{\phi_1+\phi_2}{n}}\nn\\
|w^{1/n}_1-w^{1/n}_2|^2&=&e^{\frac{\phi_1+\phi_2}{n}}\,2
\left(\cosh\left(\frac{\Delta\phi}{n}\right)-\cos\left(\frac{\Delta\tau}{nR}\right)\right).
\eea
therefore we can simplify the ratio to
\be
\frac{\langle O^\dagger(w_1,\bar{w}_1)O(w_2,\bar{w}_2)\rangle_{\Sigma_n}}{\langle O^\dagger(w_1,\bar{w}_1)O(w_2,\bar{w}_2)\rangle_{\Sigma_1}}
=\left(\frac{\cosh\left(\Delta\phi\right)
-\cos\left(\frac{\Delta\tau}{R}\right)}
{n^2\left(\cosh\left(\frac{\Delta\phi}{n}\right)
-\cos\left(\frac{\Delta\tau}{n\,R}\right)\right)}\right)^{2\Delta_O}.\label{ratioCFT}
\ee

\subsection{$\Delta/c$ corrections}\label{dcc}

In the previous subsections we assumed the factorization of the four point function in the strict large $c$ limit that is equivalent to Wick contractions. Here we will elaborate further on this limit and consider the sub-leading correction to large $c$ by including only the Virasoro conformal block from the identity operator as in (\ref{apgf}). See details in App.\ref{App:LargeC} and \cite{Fitzpatrick:2014vua} for a pedagogical review of the direct approach to Virasoro blocks.

Let us start with $n=2$ Renyi entropy and later we will generalize this analysis for higher $n$. Using the replica method, in order to evaluate $\Delta S^{(2)}_A$ for operator $O$ with (chiral) conformal dimension $\Delta_O$, we need to compute the 4-point correlator of $O$ on $\Sigma_2$. Using the conformal map $w(z)$ from the complex plane to $\Sigma_n$ we have
\be
\langle O^\dagger(w_1,\bar{w}_1)...O(w_4,\bar{w}_4)\rangle_{\Sigma_2}=\left(\prod^4_{i=1}\left|\frac{dw_i}{dz_i}\right|^{-2\Delta_O}\right)\langle O^\dagger(z_1,\bar{z}_1)...O(z_4,\bar{z}_4)\rangle_{\Sigma_1}.\label{4ptMap}
\ee
Focus now on the four point function on $\Sigma_1$ and evaluate it using the Virasoro decomposition of the conformal block. More precisely, we can rewrite it in terms of the three-point functions by inserting, after the second operator, the identity that is a sum over projectors to all states corresponding to the operators of the theory and their descendants. Performing the computation this way requires specific details of the CFT and is very complicated. However, in the limit that we are interested in, the large central charge but finite $\Delta^2_O/c$, the contribution to the four-point function of the primaries in a two-dimensional CFT\footnote{We assume that we can take the large $c$ limit.} from the identity is given by \cite{Fitzpatrick:2014vua}
\bea
\langle O^\dagger(z_1,\bar{z}_1)...O(z_4,\bar{z}_4)\rangle_{\Sigma_1}\simeq \langle O^\dagger(z_1,\bar{z}_1)O(z_2,\bar{z}_2)\rangle_{\Sigma_1} \langle O^\dagger(z_3,\bar{z}_3)O(z_4,\bar{z}_4)\rangle_{\Sigma_1}\mathcal{V}_0(z)\bar{\mathcal{V}}_0(\bar{z}).
\eea
where the vacuum conformal blocks $\mathcal{V}_0(z)$ and $\bar{\mathcal{V}}_0(\bar{z})$ at large $c$ are the following functions of the conformal cross-ratios
\be
\mathcal{V}_0(z)=\exp\left(\frac{2\Delta^2_O}{c}z^2\,_2F_1(2,2;4;z)\right),\qquad \bar{\mathcal{V}}_0(\bar{z})=\exp\left(\frac{2\Delta^2_O}{c}\bar{z}^2\,_2F_1(2,2;4;\bar{z})\right)\label{Vs}
\ee
This result is sufficient to obtain $\Delta_O/c$ corrections to our late time limit $(z,\bar{z})\to(1,0)$ of the Renyi entropies. Namely, plugging back to \eqref{4ptMap}, we can write the correction to the large $c$ factorization of the four point correlator as
\be
\langle O^\dagger(w_1,\bar{w}_1)...O(w_4,\bar{w}_4)\rangle_{\Sigma_2}=\langle O^\dagger(z_1,\bar{z}_1)O(z_2,\bar{z}_2)\rangle_{\Sigma_2} \langle O^\dagger(z_3,\bar{z}_3)O(z_4,\bar{z}_4)\rangle_{\Sigma_2}\mathcal{V}_0(z)\bar{\mathcal{V}}_0(\bar{z})+...
\ee
where ellipsis stand for contributions from operators with higher conformal dimensions.\\
Recall, that at late time  $(z,\bar{z})\to (1,0)$ and $l\ll t(\ll L)$, we find from (\ref{lik})
\be
1-z\simeq \left(\frac{\epsilon}{2t}\right)^2.\label{LTz}
\ee
After writing the hypergeometric functions in terms of the logarithm
\be
\frac{2\Delta^2_O}{c}\,z^2\,_2F_1(2,2;4;z)=-\frac{24\Delta^2_O}{c}-\frac{12\Delta^2_O}{c}\frac{2-z}{z}\log(1-z)
\ee
one can check that at late time the contribution from the $\bar{\mathcal{V}}_0(0)$ is just $1$. Finally inserting \eqref{LTz} gives us the scaling of $\Delta S^{(2)}_A$ at late time
\be
\Delta S^{(2)}_A=4\Delta_O\left(1-\frac{6\Delta_O}{c}\right)\log\left(\frac{2t}{\epsilon}\right)+\frac{24\Delta^2_O}{c}+...
\ee
As we can see the correction from the vacuum conformal block only changes the coefficient of the logarithmic growth for the late time second Renyi entropy.\\
The above analysis can be generalized to a $2n$-point functions at large $c$ (see App.\ref{App:LargeC}) . The correlator can be written as
\be
\langle O^\dagger(w_1,\bar{w}_1)...O(w_{2n},\bar{w}_{2n})\rangle_{\Sigma_2}=\left(\prod^{n}_{i=1}\langle O^\dagger(w_{2i-1},\bar{w}_{2i-1})O^\dagger(w_{2i},\bar{w}_{2i})\rangle_{\Sigma_2}\right) \mathcal{V}_0(Z_{2n})\bar{\mathcal{V}}_0(\bar{Z}_{2n})\label{2nLC}
\ee
where the conformal blocks are again given by \eqref{Vs} but the cross-ratios are in  $\{z_1,z_2,z_{2n-1},z_{2n}\}$
\be
Z_{2n}=\frac{(z_1-z_2)(z_{2n-1}-z_{2n})}{(z_1-z_{2n-1})(z_2-z_{2n})},\qquad \bar{Z}_{2n}=\frac{(\bar{z}_1-\bar{z}_2)(\bar{z}_{2n-1}-\bar{z}_{2n})}{(\bar{z}_1-\bar{z}_{2n-1})(\bar{z}_2-\bar{z}_{2n})}
\ee
Using the late time scaling of $z's$ derived in \cite{He:2014mwa} we can show that at late time
\be
1-Z_{2n}\simeq \left(\frac{\epsilon}{n\sin\left(\frac{\pi}{n}\right)\,t}\right)^2, \qquad \bar{Z}_{2n}\to 0,
\ee
and combining all together we have the corrected $n$-th Renyi entropy
\be
\Delta S^{(n)}_A=\frac{2n\Delta_O}{n-1}\left(1-\frac{12\Delta_O}{n\,c}\right)\log\left(\frac{n\sin\left(\frac{\pi}{n}\right)t}{\epsilon}\right)+\frac{24\Delta^2_O}{c(n-1)}+...
\ee
This clearly shows that the logarithmic behavior at late time persists for all the Renyi entropies $n\ge 2$ once we include the contribution from the identity.

Note that again $n\to 1$ limit is singular and is not reliable. This suggests that, at large $c$, the problem of taking the von-Neumann entropy limit $n\to 1$ is not resolved by perturbative corrections with respect to $\Delta_O/c$ and thus we need to take into account non-perturbative corrections.

\section{Holographic Analysis of Renyi Entanglement Entropy}

The AdS/CFT duality allows us to compute observables in conformal field theory from dual string theory. In the classical gravity regime of strings, that usually corresponds to a particular corner of the CFT parameter space, observables like correlation functions of primary operators can be computed by evaluating the supergravity action on the classical solution for the field dual to the operator under consideration \cite{GKP}.

In our CFT analysis of (Renyi) entanglement entropies, we tried to be as general as possible and considered a large class of field theories in two dimensions at large $c$ or higher dimensions at large $N$. In two dimensions, our main object in the replica construction of the $n$-th Renyi entropy was the $2n$-point function of operators $O(w_i,\bar{w}_i)$ on the $n$ copies of the complex plane $(\Sigma_n)$. In principle, if we knew the holographic dictionary for all the 2d CFTs on $\Sigma_n$, we could compute the $2n$-point function holographically using e.g. Witten diagrams. This is of course far from reality and we will be able to perform only a modest step towards this direction. Namely, we will approximate the $2n$-point function on $\Sigma_n$ by a product of two-point functions computed using geodesic approximation in an Euclidean AdS topological black hole by employing the coordinate transformation found in \cite{CHM}. This corresponds to the leading term in the large $N$ limit. As we learned from the field theoretic analysis, there are subtle problems of the leading term large $N$ approximation in the late time limit and $n\to 1$. In this section, we will ignore these issues and address them again in the final section.

The topological black hole, as we will review below, has an asymptotic boundary given by $S^1\times H^{d-1}$ where $H^{d-1}$ is a $d-1$-hyperbolic plane (Euclidean $AdS_{d-1}$). We can identify it with $\Sigma_n$ up-to a conformal factor \cite{CHM}. Then, similarly to \cite{Balasubramanian:1999zv}, we assume that the two-point functions of the bulk fields $\Phi_\Delta$ dual to operators $O(x)$ with (sufficiently large) dimension $\Delta$ can be computed semi-classically
\be
\langle\Phi_\Delta(x)\Phi_\Delta(x')\rangle\sim e^{-\frac{\Delta}{R}\,L(\gamma) }\label{PropGeo}
\ee
where $\gamma$ is the geodesic between the operator's insertion points that we schematically denoted $x$ and $x'$. Our approximation should be valid for a general class of CFTs at large central charge $c$. We further refer to \cite{Louko:2000tp} for a discussions on validity and subtleties of \eqref{PropGeo}.\\
In formulas below we will not use $\Phi_\Delta$'s anymore since they don't play any role in our general derivations. Instead, we identify the holographic result for the two point function on $\Sigma_n$ with the exponent of the geodesic length as
\be
\langle O^\dagger(x)O(x')\rangle_{hol}\sim e^{-\frac{2\Delta_O}{R}\,L^{(n)}}
\ee
Let us now review a few basic facts about topological black holes and proceed with computation of the geodesic lengths.

\subsection{Topological Black Holes in AdS}

The Euclidean topological black hole in $AdS_{d+1}$ is given by metric \cite{Emparan:1999gf}
\be
ds^2=f(r)d\tau^2+\frac{dr^2}{f(r)}+r^2\frac{d\zeta^2+dx_idx_i}{\zeta^2}
\ee
where
\be
f(r)=-1-\frac{\mu}{r^{d-2}}+\frac{r^2}{R^2}
\ee
It will be convenient to redefine the hyperbolic coordinate $\phi=\log\zeta$ so that the metric becomes
\be
ds^2=f(r)d\tau^2+\frac{dr^2}{f(r)}+r^2d\phi^2+r^2e^{-2\phi}dx^2_i\label{MET}
\ee
It is an example of an asymptotically $AdS$ black hole with a negatively curved horizon. The name "topological" comes from the fact that in four dimensions one can use the isometry of the hyperbolic plane $H^2$ and obtain a black hole with horizon of an arbitrary topology. Similar properties of these solutions can be demonstrated in higher dimensions but we will not need or use them here.

Topological black holes have a temperature that depends on $d$ and the horizon radius $r_+$
\be
\beta^{-1}=T_{H}=\frac{d\, r^2_+-(d-2)R^2}{4\pi R^2 r_+}
\ee
and is related to the period of the $\tau$ coordinate of the dual CFT. In our setup, we will take the period of $\tau$ to be $2\pi n R$ with integer $n$ such that
\be
\beta=2\pi n R
\ee
and as a result, the boundary will consist of $CFT_d$ on $\Sigma_n$. From this condition, we can also find $\mu$ as a function of $n$ and $d$. It is given by
\be
|\mu(d,n)|=R^{d-2}\left(\frac{1}{d\,n}+\sqrt{1-\frac{2}{d}
+\frac{1}{d^2\,n^2}}\right)^{d-2}\left(1-\left(\frac{1}{d\,n}+ \sqrt{1-\frac{2}{d}+\frac{1}{d^2\,n^2}}\right)^2\right),
\ee
and, since we will set $x_i$ to constant in \eqref{MET}, it will be our main parameter capturing the dependence on the dimensions and periodicity $n$. For clarity of our formulas we define more compactly
\be
|\mu(d,n)|\equiv R^{d-2}F_{(d,n)}\label{Fdn}
\ee
It is important to notice that for $n=1$, $\mu(d,1)$ vanishes. The opposite limit of $n\to\infty$ corresponds to the extremal black hole with a degenerate horizon.

Last but not least, in order to keep contact with the CFT results and conventions of \cite{Nozaki:2014hna}, we recall that for $\mu=0$ and $x_i=\text{const}$, metric \eqref{MET} can be mapped into $AdS_{d+1}$ in Poincare coordinates
\be
ds^2=R^2\f{dT^2+dY^2+dx_i^2}{Z^2},
\ee
by setting
\bea
Y-iT&=&\sqrt{1-\frac{R^2}{r^2}}e^{\phi-i\tau/R},　\nn\\
Y+iT&=&\sqrt{1-\frac{R^2}{r^2}}e^{\phi+i\tau/R}, \nn\\
Z&=&\frac{R}{r}e^{\phi}. \label{map}
\eea
By taking the asymptotic limit of this map we can find the relation to the polar coordinates on the plane $w=r\,e^{i\theta}$
\be
r=e^{\phi},\qquad \theta=\frac{\tau}{R}.
\ee
It is clear from this map that the space $(T,Y,x)$ precisely describe the $n$-sheeted space
$\Sigma_n$ as the periodicity of $\tau$ is $2\pi nR$.

This brief summary of the topological black hole should suffice for our further analysis and we will proceed directly with computation of the geodesic length in background \eqref{MET}. More details and further references can be found in \cite{Emparan:1999gf}.

\subsection{Geodesic length}

Let us start with the general metric \eqref{MET} and set all the $x_i=const$. As a result, the dependence on $d$ and the periodicity $n$ is fully captured by $|\mu(d,n)|$. We will compute the length of the geodesic $\left(\tau(r),\phi(r)\right)$ between the end-points $(\tau_i,\phi_i)$ and $(\tau_f,\phi_f)$ located at the boundary $r=r_\Lambda$. The geodesic extends into the bulk and has a turning point at $r=r_*$. This way, the length is twice the length from the boundary to the turning point and it is given by
\be
\frac{L}{2}=\int^{r_\Lambda}_{r_*}\sqrt{f(r)\tau'^2+\frac{1}{f(r)}+r^2\phi'^2}\,dr\equiv\int^{r_\Lambda}_{r_*}\LL\,dr
\ee
The functional neither depends on $\tau$ nor $\phi$ so the equations of motion are just the two conservation laws
\bea
\frac{f(r)\tau'}{\LL}=C_1\qquad\qquad
\frac{r^2\phi'}{\LL}=C_2
\eea
where $C_{1}$ and $C_2$ are arbitrary constants. These two ordinary differential equations can be simplified to
\bea
\frac{d\tau}{dr}&=&\frac{C_1 r}{f(r)\sqrt{f(r)(r^2-C^2_2)-r^2C^2_1}}\\
\frac{d\phi}{dr}&=&\frac{C_2}{r\sqrt{f(r)(r^2-C^2_2)-r^2C^2_1}}\label{ODEs1}
\eea
We can further relate $C_1$ to $C_2$ and $r_*$. Namely, at the turning point both derivatives $\tau'$ and $\phi'$ diverge so the denominator on the right hand side of \eqref{ODEs1} must vanish at $r=r_*$. That leads to relation
\be
C^2_1=\frac{f(r_*)(r^2_*-C^2_2)}{r^2_*}
\ee
Finally, we can insert this to \eqref{ODEs1} and, since both $\tau$ and $\phi$ should be finite, we incorporate the boundary conditions at $\infty$ in a usual manner. The procedure yields our main integrals
\bea
\Delta\tau\equiv\tau_f-\tau_i&=&2\int^{\infty}_{r_*}\frac{r\sqrt{f(r_*)(r^2_*-C^2_2)}\,dr}{f(r)\sqrt{f(r)r^2_*(r^2-C^2_2)-f(r_*)r^2(r^2_*-C^2_2)}}\\
\Delta\phi\equiv\phi_f-\phi_i&=&2\int^{\infty}_{r_*}\frac{C_2r_*\,dr}{r\sqrt{f(r)r^2_*(r^2-C^2_2)-f(r_*)r^2(r^2_*-C^2_2)}},\label{INT}
\eea
as well as the the length
\be
\frac{L}{2}=\int^{r_\Lambda}_{r_*}\frac{r_*\,r\,dr}{\sqrt{f(r)r^2_*(r^2-C^2_2)-f(r_*)r^2(r^2_*-C^2_2)}}\label{Length}
\ee
From this point we have to proceed with specific values of $d$ and $n$. \\

\subsection{Analysis in AdS$_3/$CFT$_2$}

We will begin with the simplest case that can be solved analytically, the $d=2$ that corresponds to the $AdS_3/CFT_2$ setup. Strictly speaking, there are no topological black holes in $d=2$ but we can always consider metric \eqref{MET} with $d=2$ and identify integer $n$ of the periodicity with the number of copies of the dual $CFT$. For $d=2$ we have
\be
\mu=\frac{1}{n^2}-1,\qquad \qquad r_+=\frac{R}{n}
\ee
Integrals \eqref{INT} can be performed analytically and are equal to
\bea
\frac{\Delta\tau}{R}&=&2n\arctan\left(\frac{R}{nr_*}\sqrt{\frac{r^2_*-C^2_2}{r^2_*-r^2_+}}\right)\nn\\
\Delta\phi&=&2n\,\arctanh\left(\frac{RC_2}{n\,r^2_*}\right)\label{d2Int}
\eea
Similarly, for large cut-off $r_\Lambda$, the length is given by
\be
\frac{L}{2R}=\int^{r_{\Lambda}}_{r_*}\frac{r\,dr}{\sqrt{(r^2-r^2_*)(r^2-\frac{r^2_+C^2_2}{r^2_*})}}\simeq\log\left(\frac{2r_\Lambda}{\sqrt{r^2_*-\frac{R^2C^2_2}{n^2r^2_*}}}\right)
\ee
We can easily invert \eqref{d2Int} and find $C_2$ and $r_*$ as functions of $\Delta\phi$ and $\Delta\tau$. Inserting them to the length gives
\be
\frac{L}{2R}=\log\left(\frac{nr_\Lambda}{R}\sqrt{2\cosh\left(\frac{\Delta\phi}{n}\right)-2\cos\left(\frac{\Delta\tau}{nR}\right)}\right)
\ee

Now we will use the standard semi-classical prescription for computing the propagator in a CFT using holography. Namely, the two-point function of operators with (large) dimension $\Delta$ is equal to exponent of the action of massive particle with mass $m\sim \Delta$ evaluated on particle's geodesic with boundary conditions determined by the insertion points of operators in the CFT. Assuming that such approximation holds for our setup as well, the two-point function of operators with dimension $\Delta=2\Delta_O$ in $n$ copies of $CFT_2$ ($\Sigma_n$) can be approximated by the exponent of the geodesic length in the topological black hole
\be
e^{-\frac{2\Delta_O}{R} L^{(n)}}=\left(\frac{1}{\frac{2n^2\,r^2_\Lambda}{R^2}\left(\cosh\left(\frac{\Delta\phi}{n}\right)-\cos\left(\frac{\Delta\tau}{n\,R}\right)\right)}\right)^{2\Delta_O}
\ee
By construction, when computing $\Delta S^{(n)}_A$ at late time, we will be mostly interested in the ratio of the two-point function on $\Sigma_n$ to the two-point function on $\Sigma_1$. The relevant object will then be
\be
e^{-\frac{2\Delta_O}{R}\left(L^{(n)}-L^{(1)}\right)}=\left(\frac{\cosh\left(\Delta\phi\right)-\cos\left(\frac{\Delta\tau}{R}\right)}{n^2\left(\cosh\left(\frac{\Delta\phi}{n}\right)-\cos\left(\frac{\Delta\tau}{n\,R}\right)\right)}\right)^{2\Delta_O}
\label{ratiod2}
\ee
One can verify that this ratio perfectly matches the $CFT_2$ result \eqref{ratioCFT} computed using a conformal map.

Our goal is to determine the late time behavior of the ratio of the $2n$-point function on $\Sigma_n$ to the n-th power of the two-point function on $\Sigma_1$.  We approximate the correlator on $\Sigma_n$ with "free field Wick contractions" using geodesics in the topological black hole. The geodesics are stretched between the insertion points of the operators at the boundary that are \cite{Nozaki:2014hna} $w_i=\exp(\phi_i+i\tau_i/R)$, $(i=1,...,2n)$, and points $w_i$ are given in \eqref{wco},\eqref{wcor}.
It is then a simple exercise to show that
\bea
\cos\left(\frac{\tau_1-\tau_2}{R}\right)&=&\frac{l^2-\epsilon^2-t^2}{\sqrt{(l^2-t^2+\epsilon^2)^2+4\epsilon^2t^2}}\nn\\
\cosh\left(\phi_1-\phi_2\right)&=&\frac{l^2+\epsilon^2-t^2}{\sqrt{(l^2-t^2+\epsilon^2)^2+4\epsilon^2t^2}}
\eea
and in the late time limit $(t\gg l \gg \epsilon)$ we have
\bea
\cos\frac{\Delta\tau}{R}&=&-1+O(t^{-4})\\
\cosh\Delta\phi&=&-1+\frac{2\epsilon^2}{t^2}+O(t^{-4})
\eea
In terms of $\tau$ and $\phi$, that corresponds to
\be
\frac{\Delta\tau}{R}=\pi+O(t^{-2}),\qquad \Delta\phi=i\left(\pi-\frac{2\epsilon}{t}\right)+O(t^{-2})
\ee
In our solutions \eqref{d2Int}, the late time limit can be extracted by first analytically continuing $C_2\to i |C_2|$ and then taking $r_*$ very large while keeping $\beta=\frac{R |C_2|}{r^2_*}$ fixed but also large. Since all the transverse directions in \eqref{MET} are fixed to constant, we take this as the universal late time limit in any dimensions $d$.\\
In the late time limit the ratio \eqref{ratiod2} becomes
\be
e^{-\frac{2\Delta_O}{R}\left(L^{(n)}-L^{(1)}\right)}=\left(\frac{\epsilon}{n\sin\left(\frac{\pi}{n}\right)t}\right)^{2\Delta_O}\label{LTratio}
\ee
Then, the leading contribution (the shortest length) to the $2n$-point correlator comes from two-possible Wick contractions (remember Fig.3). Namely, the first in which we contract the pairs of operators on the same sheet and the second, when the operator on the $i$-th sheet is contracted with the operator on sheet $i+1$ ( operator on the $n$-th sheet is contracted with the operator on sheet $i=1$). As a result, using the holographic ratio \eqref{LTratio}, we obtain the late time limit of the Renyi entropy
\be
\Delta S^{(n)}_A\simeq\frac{1}{1-n}\log\left[2\left(\frac{\epsilon}{n\sin\left(\frac{\pi}{n}\right)t}
\right)^{2n\Delta_O}\right]. \label{holren}
\ee
More explicitly, for example for $n=2$ in $d=2$ we have
\be
\Delta S^{(2)}_A=4\Delta_O\log\left(\frac{2t}{\epsilon}\right)-\log 2.
\ee

Our holographic results (\ref{holren}) perfectly agree with our 2d CFT results (\ref{DS3}) in the large $c$ limit. This confirms our large $c$ approximation studied in section 4 corresponds to the dual gravity calculation with the geodesic approximation, which ignores all backreactions of the massive particle in the AdS.

The geodesic approximation regards the $2n$ point function as $n$ disconnected two point functions. As usual, this can be naively justified in the large $N$ limit. However, our previous CFT arguments suggests that this approximation breaks down when $t$ is very large ($t\to\infty$) as in (\ref{sppqq}) or when we take the von-Neumann entropy $n\to 1$ limit. In both cases, it is expected that we need to take into account non-perturbative corrections about the large $N$ limit.

\subsection{Analysis in Higher Dimensions}

For higher $d$ the integrals become more complicated and the analytical answer seems very hard to obtain (inverting $C_2$ and $r_*$). However we can still extract the late time answer ($t>> l$ but still smaller than $\infty$) for $\Delta S^{(n)}_A$. Here we briefly describe the procedure and state the main results but more details can be found in App. \ref{GeoHd}.

Analyzing the geodesic lengths in $d\ge 2$ (see App.\ref{GeoHd} for details) one can check that at late time $\Delta S^{(n)}_A$ has the same divergent behavior
\be
\Delta S^{(n)}_A\simeq\frac{4n\Delta}{d(n-1)}\log\left(\frac{F_{(d,n)}\,t}{2\,\epsilon}\right)+C_{(n,d)}-\frac{1}{n-1}\log 2\label{LTd}
\ee
where $C_{(n,d)}$ is some non-universal constant and $F_{(d,n)}$ was defined in \eqref{Fdn}.
Note that higher Renyi entropies all behave similarly and for example in the extremal limit we get the min-entropy
\be
\Delta S^{(\infty)}_A\simeq \frac{4\Delta}{d}\log\left(\frac{t}{\epsilon}\right)
\ee

Again, we expect that our geodesic approximation may break down when $t$ is very large as in the 2d CFT case. It is also clear that for any $d$, in the limit of $n\to 1$ we cannot trust the holographic computation.  In both cases, we expected that non-perturbative corrections about the large $N$ limit will become important.

\section{Holographic Analysis of von-Neumann Entanglement Entropy}

In the previous section, we holographically evaluated the Renyi entanglement entropy in the Replica method by employing the heavy particle geodesic approximation of two point functions in AdS/CFT. However, this approximation, which keeps the only leading term in the large $N$ expansion, is valid for $n>1$ and
we cannot obtain any reliable result for the von-Neumann entanglement entropy $n=1$. Obviously, one way to deal with this problem is to take into account the backreactions in our topological black hole analysis by computing directly the Witten diagrams for the holographic $2n$ point functions as in \cite{GKP}. Unfortunately this requires rather complicated analysis and we do not want to pursuit it here. Moreover, our previous large $c$ analysis in 2d CFT suggests that we may need to incorporate non-perturbative corrections of quantum gravity to get sensible results in this limit.

Therefore we would like to consider some other holographic calculations for $n=1$. Indeed, the holographic entanglement entropy \cite{RT} offers another method to compute von-Neumann entanglement entropy without referring to the Renyi entropy. In this calculation, we prepare a (time-dependent) gravity background dual to the excited state defined by a local operator in CFT and compute the extremal surface area.

Actually we would like to point out that the holographic entanglement entropy for the excited state by local operator coincides with the calculation done in the previous paper \cite{NNT}. In \cite{NNT}, the gravity dual is obtained by a heavy falling particle (with mass $m$) in AdS and the holographic entanglement entropy has been computed by taking into account gravity backreactions analytically. Thus this backreacted geometry offers the gravity dual of the excited state by a local operator with conformal dimension $\Delta \simeq mR$. Indeed, we can confirm that
the holographic stress energy tensor in this gravity background precisely reproduces (\ref{energyd}) for the AdS$_3/$CFT$_2$ setup. For a large mass $m$ such that $\Delta \sim O(N^2)=O(c)$ we will be able to regard it as a version of local quench as it causes a very large excitation as argued in \cite{NNT}. We are interested in the opposite case $\Delta \ll c$.

Let us focus on the AdS$_3/$CFT$_2$ case so that we get analytical results.
In the construction of \cite{NNT}, the regularization parameter\footnote{Please note that in the paper \cite{NNT} this parameter $\ep$ is written as $\ap$.} $\ep$ appears as a parameter of the trajectory of the falling massive particle: $z=\s{t^2+\ep^2}$ in the Poincare AdS$_3$ space
$ds^2=R^2\left(\f{dz^2-dt^2+dx^2}{z^2}\right)$. Then, when the subsystem $A$ is given by a half line $x>0$ and we insert the local operator at $(t,x)=(0,-l)$, the final result of time evolution of holographic entanglement entropy looks as follows \cite{NNT}:
\ba
&& \Delta S^{(1)}_A=0 \ \ (t<l),  \no
&& \Delta S^{(1)}_A\simeq \f{c}{6}\log \f{t}{\ep}+\f{c}{6}\log \f{\Delta}{c} \ \
\left(t\gg l\right).  \label{lqee}
\ea
Firstly we note that the slope of the logarithmic time evolution is given by a constant of the CFT i.e. the central charge $c$. However, if we take $\Delta \to 0$ limit, we get $\Delta S^{(1)}_A=0$ for any time because the logarithmic region is pushed into the infinite late time and disappears, as expected.

Even though the logarithmic behavior occurs also in our previous holographic analysis of Renyi entanglement entropy in terms of $2n$ point functions, the coefficient of the logarithmic term $\log t/\ep$ is given by $\f{2n\Delta}{n-1}$ (\ref{holren}).  This gets divergent in the von-Neumann limit $n\to 1$, where the approximation by the large $N$ limit breaks down as we mentioned. It is possible to speculate that the regularization of this divergence comes from summing over non-perturbative corrections of the large $N$ expansion and that this gives the finite gradient proportional to the central charge $c$ as we will discuss in the next section. It will be a very important future problem to study this issue closely.

\section{Summary and Discussions}

In this paper we studied the Renyi and von-Neumann entanglement entropy of excited states produced by local operators in large $N$ (or equally large $c$) CFTs with holographic duals. The Renyi entanglement entropy growth $\Delta S^{(n)}_A$ can be computed from a $2n$ point correlation function. We did both field theoretic and holographic analysis in various dimensions. We mainly chose the subsystem $A$ to be simply the half space.

Firstly, we studied these quantities in free CFTs.  Especially we analyzed the scalar operator in a free $U(N)$ Yang-Mills theory in four dimensions. We found that the Renyi and von-Neumann entanglement entropy both grow monotonically in time and eventually approaches to finite constants $\Delta S^{(n)}_A$. We computed them analytically and exactly. We also found that the Renyi entanglement entropy $n>1$ always scales like $\Delta S^{(n)}_A\simeq \f{Jn-1}{n-1}\log 2$ as in (\ref{free}), where $J$ is the number of scalar fields in the local operator or equally its conformal dimension. However, in the von-Neumann entropy limit $n\to 1$, we immediately find that this gets divergent. This is because the large $N$ expansion breaks down and we have to sum over many terms, where we cannot ignore the connected diagrams even for the dominant contributions. In other words, some of subleading terms scale as some powers of $N^{1-n}$, which can be ignored only when $n>1$. By taking these contributions into account, we find that the von-Neumann entanglement entropy scales like $\Delta S^{(1)}_A\sim J\log N$, which has the logarithmic enhancement. This result is consistent with the intuitive argument that since the scalar field is a $N\times N$ matrix, each scalar field should produce $\sim\log N^2$ entanglement entropy. Our analysis reveals an intriguing phenomenon where the Renyi and von-Neumann entanglement entropy behave differently. As a consequence, we have to be very careful when we take the von-Neumann entropy limit $n\to 1$ for excited states.

Next, as another main result of this paper, we analyzed these entropies in strongly coupled and large $N$ CFTs so that they have classical gravity duals via the AdS/CFT. We did this analysis for two dimensional CFTs in the large $c$ limit. Then, we  pursued two holographic computations: one was the geodesic approximation of correlation functions in the replica calculations of Renyi entanglement entropy. The other was the analysis of von-Neumann entanglement entropy using the holographic entanglement entropy, which is essentially same as the earlier work \cite{NNT}. Below we would like to summarize main part of our results schematically.

Generally within our analysis, being weakly coupled or strongly coupled, the growth of entanglement entropy $\Delta S^{(n)}_A$ behaves as follows (assuming the limit $t\gg l\gg \ep$, where we can suppress $l$ dependence):
\ba
\Delta S^{(n)}_A&=&-\f{1}{n-1}\log{\f{\langle O^{\dagger}O
\!\ddd\! O^{\dagger} O \rangle_{\Sigma_n}}{\left(\langle O^{\dagger} O
\rangle_{\Sigma_1}\right)^n}} \no
&\simeq & -\f{1}{n-1}\log\left[\f{1}{D_n}+\mu_n\cdot \left(\f{\ep}{t}\right)^{\nu_n}\right].
\label{ddewss}
\ea
 Here, $D_n(>0),\nu_n(>0)$ and $\mu_n$ are $n$ dependent constants. Since the $2n$ point function is reduced to the two point function at $n=1$, we obtain the relation
\be
\f{1}{D_1}+\mu_1\cdot \left(\f{\ep}{t}\right)^{\nu_1}=1.   \label{relqpa}
\ee

In our free field theory example of four dimensional $U(N)$ scalar field, for the Renyi entanglement entropy $n\geq 2$, we found
\be
D_n=2^{Jn-1}+O(N^{-2}),   \label{freedec}
\ee
while $\mu_n=O(2^{-Jn})$ and $\nu_n=O(1)$.
If we take the von-Neumann limit $n\to 1$ we cannot ignore subleading terms of large $N$ expansion in (\ref{freedec}) as we mentioned. However, once we take this into account, the late time limit $t\to \infty$ is smoothly taken and we obtain a finite entropy. A similar behavior is true for the rational CFTs in two dimension, where
$D_n$ is given by the quantum dimension $d_O$ of the primary operator $O$ as $D_n=(d_O)^{n-1}$ \cite{He:2014mwa} and therefore $\Delta S^{(n)}_A$ approach the same value
$\log d_O$ for any $n$ in the late time limit $t\to \infty$. It is also helpful to note that in these examples, there is no time range where the second term in (\ref{ddewss}), which is time-dependent, gets dominant over the first term $1/D_n$. In other words, we cannot find the logarithmic grows of $\Delta S^{(n)}_A$ in the above examples.

On the other hand, our holographic analysis of strongly coupled large $N$ CFTs in $d$ dimensions lead to following behavior for an operator with the conformal dimension $\Delta$:
\be
\nu_n\simeq \f{4n\Delta}{d}+O\left(\f{\Delta^2}{c}\right), \label{renholg}
\ee
where $c\sim N^2$ is the central charge of each CFT. We reproduced the same result for $d=2$ from field theoretic calculations. The fact that the constant part, i.e. $1/D_n$ term, is missing in the perturbative large $N$ expansions, suggest that it behaves like
\be
D_n\sim e^{b_n\cdot N^{a_n}},
\ee
where $a_n$ and $b_n$ are positive constants. This of course corresponds to a non-perturbative contribution in the large $N$ expansion. In our holographic and field theoretic arguments, it was very difficult to compute this non-perturbative term $D_n$. However, since we expect that it is very large, we find that there is a long time period where $\Delta S^{(n)}_A$ grows logarithmically with time as
\be
\Delta S^{(n)}_A\simeq \f{\nu_n}{n-1}\log\f{t}{\ep}, \label{eelogh}
\ee
 as in (\ref{sppqq}). It is intriguing to note that this logarithmic growth is peculiar to strongly coupled large $N$ CFTs.  At late time limit
$t\to \infty$, $\Delta S^{(n)}_A$ approaches\footnote{However, notice that from our computations in this paper we cannot completely deny the possibility that $D_n$ is infinitely large such that $\f{1}{D_n}=0$ precisely, though physically this may look unusual. If this happens, even in the late time limit $t\to \infty$, the Renyi entanglement entropy grows logarithmically and does not approach to a constant. On the other hand, our previous rough argument in the footnote 5 and its generalization to any $n$ following \cite{He:2014mwa}, give the estimation $D_n\gtrsim e^{p_n(n-1)c}$, where $p_n$ is a positive $O(1)$ constant. If this is true, both Renyi and von-Neumann entanglement entropy approaches finite constants $\Delta S^{(n)}_A\gtrsim O(c)$ in the late time limit $t\to \infty$. It is very important future problem to examine this issue more closely.} to a constant $\f{1}{n-1}\log D_n$.

The von-Neumann limit $n\to 1$ is more subtle in strongly coupled large $N$ CFTs than that in free field CFTs. The leading order results (\ref{renholg}) and (\ref{eelogh}) already tell us that
$\Delta S^{(n)}_A$ gets divergent at $n=1$ and thus this says that
the large $N$ expansion will break down again. On the other hand, our holographic result (\ref{lqee}) based on the holographic entanglement entropy \cite{NNT} implies the following behavior
\be
\nu_n=(n-1)\cdot \nu+O\left((n-1)^2\right),
\ee
where $\nu$ should satisfy
\be
\mu_1\nu=\f{c}{6}.
\ee
 Moreover, the relation (\ref{relqpa}) leads to
\be
\f{1}{D_1}+\mu_1=1.
\ee
These suggest that $\nu\sim O(c)$ and this is possible only if we assume non-perturbative terms in (\ref{renholg}).

This von-Neumann entanglement entropy grows logarithmically $\Delta S^{(1)}_A\simeq \f{c}{6}\log\f{t}{\ep}$ even in the late time limit. This looks similar to the local quantum quenches \cite{cal}. However, notice the difference that in our setup we consider a single local operator excitation and that it is much simpler than the one for local quenches, which are triggered by a sudden change of Hamiltonian at a particular point and which produce infinitely many operators. In this way, the results for the strongly coupled large $N$ CFTs looks different from those for free CFTs. Therefore it is an important future problem to confirm our prediction for strongly coupled CFTs and understand both $n\to 1$ limit and the late time limit $t\to\infty$ in more details.

\subsection*{Acknowledgements}

We would like to thank Sumit Das, Roberto Emparan, Stefano Giusto, Vishnu Jejjala, Tokiro Numasawa, Hesam Soltanpanahi, Rodolfo Russo and Kento Watanabe for comments and useful discussions and especially to Sylvain Ribault for detailed explanations of a large central charge limit of 2d CFTs. PC and MN are supported by JSPS fellowship. TT is supported by JSPS Grant-in-Aid for Scientific Research (B) No.25287058 and by JSPS Grant-in-Aid for Challenging Exploratory Research No.24654057. TT is also supported by World Premier International Research Center Initiative (WPI Initiative) from the Japan Ministry of Education, Culture, Sports, Science and Technology (MEXT).

\begin{appendix}

\section{$2n$-point correlators at large $c$}\label{App:LargeC}

In this appendix we briefly summarize the analysis of \cite{Fitzpatrick:2014vua} on direct computation of the vacuum Virasoro conformal blocks at large central charge. We also generalize the large $c$ results to $2n$-point functions and outline the derivation of formula \eqref{2nLC} from the main text.\\
Let us start with the four-point correlator of primary operators in a two-dimensional CFT at large central charge. We can insert to the correlator the identity, that is, a sum over projectors on states corresponding to all the operators of the theory and their descendants. Schematically we have
\be
\langle O^\dagger(z_1,\bar{z}_1)...O(z_4,\bar{z}_4)\rangle_{\Sigma_1}
=\sum_{\alpha_h,\bar{\alpha}_{\bar{h}}}\langle O^\dagger(z_1,\bar{z}_1)O(z_2,\bar{z}_2)
\ket{\alpha_h,\bar{\alpha}_{\bar{h}}}
\bra{\bar{\alpha}_{\bar{h}},\alpha_h}
O^\dagger(z_3,\bar{z}_3)O(z_4,\bar{z}_4)\rangle_{\Sigma_1}
\ee
where $\alpha_h$ ( $\bar{\alpha}_{\bar{h}}$ ) stands for all operators of dimension $h$ $(\bar{h})$ and their descendants.\\
In \cite{Fitzpatrick:2014vua} authors derived the contribution from the identity operator  (and its descendants) $\ket{\alpha_0}$ in the limit of non-vanishing $\Delta^2_O/c$. A convenient basis for such computations consist of $k$-graviton states \footnote{Projection operators are tensor products of the holomorphic and anti-holomorphic parts so we treat explicitly only the holomorphic contribution and simply multiply the answer by the anti-holomorphic counterpart}
\be
\ket{\alpha_0}=\frac{L^{k_1}_{-m_1}...L^{k_p}_{-m_p}\ket{0}}{\sqrt{\mathcal{N}_{\{m_i,k_i\}}}}
\ee
where $k_1+...+k_p=k$, $m_1>...>m_p$, and the creation operators $L_{-m}$, with $m\ge2$, are the generators of the Virasoro algebra
\bea
[L_m,L_{n}]&=&(m-n)L_{m+n}+\frac{c}{12}m(m^2-1)\delta_{m,-n}\label{Vir}
\eea
First, one can easily check that $k$-graviton states are orthogonal only at large $c$; the overlap between states with different numbers of $L_{-m}$'s can be neglected at large c. Moreover, applying \eqref{Vir}, one can also show that the leading large $c$ contribution to the norm of the state with $k$-gravitons  is given by
\be
\mathcal{N}_{\{m_i,k_i\}}\sim\left(\frac{c}{12}\right)^k\prod^{p}_{i=1} k_i!m^{k_i}_i(m^2_i-1)^{k_i}
\ee
As shown in  \cite{Fitzpatrick:2014vua}, using \eqref{Vir} as well as the commutation relation for generators $L_{-m}$ with primary operators
\bea
[L_{\pm m},O(z_i,\bar{z}_i)]=\left(\Delta_O(1\pm m)z^{\pm m}_i+z^{1\pm m}_i\partial_{z_i}\right)O(z_i,\bar{z}_i)\label{ComPr}
\eea
after summing over all $k$-graviton contributions we get the large $c$ part of the four-point function that comes from the identity operator
\bea
\langle O^\dagger(z_1,\bar{z}_1)...O(z_4,\bar{z}_4)\rangle_{\Sigma_1}\sim \langle O^\dagger(z_1,\bar{z}_1)O(z_2,\bar{z}_2)\rangle_{\Sigma_1} \langle O^\dagger(z_3,\bar{z}_3)O(z_4,\bar{z}_4)\rangle_{\Sigma_1}\mathcal{V}_0(z)\bar{\mathcal{V}}_0(\bar{z})
\eea
where the vacuum conformal blocks at large $c$ are given by exponentials of the single graviton contribution (see \cite{Fitzpatrick:2014vua} for detailed derivation)
\be
\mathcal{V}_0(z)=\exp\left(\frac{2\Delta^2_O}{c}z^2\,_2F_1(2,2;4;z)\right),\qquad \bar{\mathcal{V}}_0(\bar{z})=\exp\left(\frac{2\Delta^2_O}{c}\bar{z}^2\,_2F_1(2,2;4;\bar{z})\right)
\ee
where $z$ and $\bar{z}$ are the conformal cross-ratios constructed from $\{z_1,z_2,z_3,z_4\}$.\\
Their analysis can be extended for our purposes to the $2n$-point correlator of the primary fields satisfying \eqref{ComPr}. Let us first look at the contribution from a single graviton $L_{-m}$ (holomorphic part).  To keep our formulas compact we label the operators by their position $O_i\equiv O(z_i,\bar{z}_i)$. We insert the $n-1$ projection operators after each pair of operators
\be
G^{(k=1)}_{2n}\equiv\sum^{\infty}_{\{m\}=2}\mathcal{N}_{\{m\}}\langle O_{1}O_{2}L_{-m_1}\rangle\langle L_{m_1}O_3O_4L_{-m_2}\rangle...\langle L_{m_{n-2}}O_{2n-3}O_{2n-2}L_{-m_{n-1}}\rangle\langle L_{m_{n-1}}O_{2n-1}O_{2n}\rangle\label{G2n}
\ee
where the norm is given by
\be
\mathcal{N}_{\{m\}}=\left(\frac{12}{c}\right)^{n-1}\,\prod^{n-1}_{k=1}\frac{1}{m_k(m^2_k-1)}
\ee
The building blocks in \eqref{G2n} can be again computed at large $c$ using \eqref{Vir} and \eqref{ComPr}. The three-point functions are
\bea
\langle O_{i}O_jL_{-m}\rangle&=&\Delta_O\left((m-1)(z^{-m}_i+z^{-m}_j)+\frac{2}{z_{ij}}(z^{1-m}_i-z^{1-m}_j)\right)\langle O_{i}O_j\rangle\nn\\
\langle L_{m}O_{i}O_j\rangle&=&\Delta_O\left((m+1)(z^{m}_i+z^{m}_j)-\frac{2}{z_{ij}}(z^{1+m}_i-z^{1+m}_j)\right)\langle O_{i}O_j\rangle
\eea
and the four point functions with $L_n$, $n\ge 2$ and $L_{-m}$, $m\ge 2$ are explicitly given by
\bea
\langle L_n O_{i} O_{j} L_{-m}\rangle&=&\langle[L_n,O_i]O_j L_{-m}\rangle+\langle O_i[L_n,O_j] L_{-m}\rangle+\langle O_{i}O_j [L_n,L_{-m}]\rangle\nn\\
&=&\sum_{k=\{i,j\}}\left(\Delta_O(1+n)z^{n}_k+z^{1+n}_k\partial_{z_k}\right)\langle O_{i}O_jL_{-m}\rangle \nn\\
&+& \langle O_{i}O_j [L_n,L_{-m}]\rangle
\eea
Notice that the last term is equal to
\be
\langle O_{i}O_j [L_n,L_{-m}]\rangle=(n+m)\langle O_{i}O_j L_{n-m}\rangle+\frac{c}{12}n(n^2-1)\delta_{n,m}\langle O_{i}O_j\rangle
\ee
It is then clear that at large $c$
\be
\langle L_n O_{i} O_{j} L_{-m}\rangle\sim\frac{c}{12}n(n^2-1)\delta_{n,m}\langle O_{i}O_j\rangle
 \ee
Inserting this to $G^{(k=1)}_{2n}$ yields
\bea
G^{(k=1)}_{2n}&\sim&\prod^{n-1}_{i=2}\langle O_{2i-1}O_{2i}\rangle\sum^{\infty}_{\{m\}=2}\mathcal{N}_{\{m\}}\left(\prod^{n-2}_{l=1}\frac{c}{12}m_l(m^2_l-1)\delta_{m_l,m_{l+1}}\right)\langle O_{1}O_{2}L_{-m_1}\rangle\langle L_{m_{n-1}}O_{2n-1}O_{2n}\rangle\nn\\
&=&\left(\prod^{n-1}_{i=2}\langle O_{2i-1}O_{2i}\rangle\right) \frac{12}{c}\sum^{\infty}_{m_{1}=2}\frac{1}{m_{1}(m^2_{1}-1)}\langle O_{1}O_{2}L_{-m_{1}}\rangle\langle L_{m_{1}}O_{2n-1}O_{2n}\rangle\nn\\
&=&\left(\prod^{n}_{i=1}\langle O_{2i-1}O_{2i}\rangle\right) \mathcal{V}^{(k=1)}_0(Z_{2n})
\eea
where the conformal block is
\be
\mathcal{V}^{(k=1)}_0(Z_{2n})=\frac{2\Delta^2_O}{c}Z^2_{2n}\,_2F_1(2,2;4;Z_{2n})
\ee
and $Z_{2n}$ is the conformal cross-ratio constructed from $\{z_1,z_2,z_{2n-1},z_{2n}\}$. Notice, that the contribution from all the four-point functions is such that precisely cancels the $(n-2)$ norms and at the end we only have a single factor of $\Delta^2_O/c$. All the other contributions would be suppressed by higher powers of $1/c$.\\
Taking into account also the anti-holomorphic part, we have the contribution from a single graviton
\be
G^{(k=1)}_{2n}= \left(\prod^{n}_{i=1}\langle O_{2i-1}O_{2i}\rangle\right) \mathcal{V}^{(k=1)}_0(Z_{2n})\bar{\mathcal{V}}^{(k=1)}_0(\bar{Z}_{2n})+...
\ee
Now, the leading $c$ contribution from $k$-graviton states must be the one that cancels the "maximal" number of the norms with powers of $c^{k}$. This is precisely the contribution with the same structure of $L$'s on both sides of the four point functions and is proportional to the product of $\delta_{m,n}$'s. Finally, summing up all the $k$-graviton contributions and noticing that the contribution of each of the $k$-gravitons commutes with each other \cite{Fitzpatrick:2014vua} we get the exponentiation of the single graviton answer \eqref{2nLC} at large $c$.

\section{Geodesic length in $d>2$}\label{GeoHd}

For $d>2$ the integrals \eqref{INT} become much more complicated and even though for $d=3,4$ one can still integrate them to known Elliptic functions, inverting them to find $C_2$ and $r_*$ becomes a formidable task. Nevertheless, in order to extract the geodesic length at late time, we can just expand the integrals for large $r_*$  and find the answer to an arbitrary order of precision. Let us carefully go through the analysis in $d=3$.
The relevant denominator that appears in the integrals \eqref{INT} is given by
\bea
Den_3(x)=\frac{r^6_*(x-1)pol_3(x)}{R^2\,x}
\eea
where
\bea
pol_3(x)=x^4+x^3+x^2\left(\frac{|C_2|^2R^2}{r^4_*}-\frac{|\mu|R^2}{r^3_*}-\frac{|\mu||C_2|^2R^2}{r^5_*}\right)\nn\\
+x\left(\frac{|C_2|^2R^2}{r^4_*}-\frac{|\mu||C_2|^2R^2}{r^5_*}\right)-\frac{|\mu||C_2|^2R^2}{r^5_*}
\eea
 We have performed analytic continuation $C_2\to i C_2$ ($C^2_2\to -|C^2_2|$) because we want $\Delta\phi$ to be purely imaginary at late time (recall the discussion from the main text).
Now we define
\be
\beta=\frac{R |C_2|}{r^2_*},
\ee
expand the integrals
\bea
|\Delta\phi|&=&2\beta\int^{\infty}_{1}\frac{dx}{\sqrt{x(x-1)pol_3(x)}}\\
\frac{\Delta\tau}{R}&=&2\beta\int^{\infty}_1\frac{\sqrt{1+\frac{R^2}{\beta^2r^2_*}}\sqrt{1-\frac{R^2}{r^2_*}+\frac{|\mu|R^2}{r^3_*}}x^{5/2}dx}{\left(x^3-\frac{R^2}{r^2_*}x+\frac{|\mu|R^2}{r^3_*}\right)\sqrt{(x-1)pol_3(x)}}
\eea
to order $r^{-3}_*$ and integrate. The answer reads
\bea
\pi-\frac{2\epsilon}{t}=|\Delta\phi|&=&\pi-\frac{2}{\beta}+\frac{2}{3\beta^3}+\frac{2|\mu|}{r_*}-\frac{2|\mu|}{r_*\beta}+\frac{15\pi-16}{16}\frac{|\mu|^2}{r^2_*}\nn\\
&-&\frac{3|\mu|}{2r_*\beta^2}-\frac{2|\mu|^2}{\beta r^2_*}+\left(\frac{61}{12}-\frac{15\pi}{16}\right)\frac{|\mu|^3}{r^3_*}+O(\varepsilon^4)\\
\nn\\
\pi=\frac{\Delta\tau}{R}&=&\pi-\frac{2}{\beta}+\frac{2}{3\beta^3}+\frac{2|\mu|}{r_*}-\frac{2|\mu|}{r_*\beta}+\frac{15\pi-16}{16}\frac{|\mu|^2}{r^2_*}\nn\\
&-&\frac{3|\mu|}{2r_*\beta^2}+\frac{R^2}{r^2_*\beta}-\frac{2|\mu|^2}{r^2_*\beta}+\frac{R^2|\mu|}{3r^3_*}+\left(\frac{61}{12}-\frac{15\pi}{16}\right)\frac{|\mu|^3}{r^3_*}+O(\varepsilon^4)
\eea
where we denote $\varepsilon=|\mu|/r_*$. Once we subtract the two equations we get a relation between the parameters
\be
\frac{2\epsilon}{t}=\frac{R^2}{r^2_*\beta}+\frac{R^2|\mu|}{3r^3_*}
\ee

This algorithm can be repeated for any $d$ and we have verified that in $d$ dimensions the relation becomes
\be
\frac{2\epsilon}{t}\sim \frac{R^2}{r^2_*\beta}+\lambda_{d}\frac{R^2|\mu|}{r^d_*}
\ee
where $\lambda_d$ is some $O(1)$ number that depends on $d$.\\
Now comes the crucial observation: in order for the terms on the right hand side of this equation to be of the same order (as we assume in our expansion) we must have
\be
\beta\sim \alpha_d \frac{r^{d-2}_*}{|\mu|}\label{beta}
\ee
where $\alpha_d$ is some $O(1)$ number. Then
\be
\frac{2\epsilon}{t}\sim \frac{R^2|\mu|}{r^d_*}\left(\frac{1}{\alpha_d}+\lambda_d\right)\equiv c_d\frac{R^2|\mu|}{r^d_*}\label{relation}
\ee
where in d-dim we have $dim(|\mu|)=d-2$.\\
Finally, from \eqref{Length}, we get the length of the geodesic to the leading order in $\beta$
\be
\frac{L^{(n)}_{12}}{2R}\sim\log\left(\frac{2r_\Lambda}{r_*\beta}\right)
=\log\left(\frac{2r_\Lambda|\mu(n)|}{\alpha_d\, r^{d-1}_*}\right)
\ee
what after using \eqref{relation} gives the late time propagator in $d$ dimensions
\be
e^{-\frac{2\Delta_O}{R} L^{(n)}_{12}}\sim\left(\frac{\alpha_dR^{\frac{d-2}{d}}}
{\frac{2r_\Lambda}{R}|\mu|^{\frac{1}{d}}}
\left(\frac{2\epsilon}{c_dt}\right)^{\frac{1}{d}-1}\right)^{4\Delta_O}
\ee
This leads to the leading order contribution to $\Delta S^{(n)}_A$ \eqref{LTd}.

\section{Higher $J$ operators}\label{HigherJ}

In this appendix we present some details of the direct computation of $\Delta S^{(n)}_R$. We have carried them out up to length 5 but already for $J=3$ one can see the subtleties and understand the general procedure behind the computation.  For $J=3$ we expand the operator $\Tr(\phi^3)$ into left- and right- moving parts
\be
\ket{\psi_3}=\ket{\Tr\phi^3_L}\ket{0_R}+\ket{0_L}\ket{\Tr\phi^3_R}
+3\sum^{N}_{\alpha,\beta=1}\ket{\left(\phi^2_L\right)^\alpha_\beta}
\ket{\left(\phi_R\right)^\beta_\alpha}+3\sum^{N}_{\alpha,\beta=1}
\ket{\left(\phi_L\right)^\alpha_\beta}\ket{\left(\phi^2_R\right)^\beta_\alpha}
\ee
where we again write explicitly the summation over indices in traces that contain both $\phi_L$ and $\phi_R$.
The orthonormal basis should be given by
\be
\left\{\ket{0_{L}},\frac{1}{\sqrt{3N(N^2+1)}}\ket{\Tr(\phi^3_{L})},
\ket{\left(\phi_{L}\right)^\alpha_\beta},\mathcal{N}_{\alpha,\beta}
\ket{\left(\phi^2_{L}\right)^\alpha_\beta}\right\}
\ee
However, note that the scalar product for two of the last states with different indices is
\be
\left<\left(\phi^2_{L}\right)^a_b\right|\left.\left(\phi^2_{L}\right)^c_d\right>=N\,\delta^a_d\delta^c_b+\delta^a_b\delta^c_d\label{ort}
\ee
so these states are orthogonal only at large $N$. Let us then focus on $n\ge 2$ and proceed assuming large N (for finite N we should probably work with Schur polynomials instead of traces, see \cite{Shurs} for the appropriate technology for $U(N)$ and other classical gauge groups) with basis
\be
\left\{\ket{0_{L}},\frac{1}{\sqrt{3N^3}}\ket{\Tr(\phi^3_{L})},\ket{\left(\phi_{L}\right)^\alpha_\beta},\frac{1}{\sqrt{N}}\ket{\left(\phi^2_{L}\right)^\alpha_\beta}\right\}
\ee
After normalization the reduced density matrix becomes
\be
\rho_R=\text{diag}\left(\frac{1}{8},\frac{1}{8},\frac{9\times\vec{1}}{24N^2},\frac{9\times\vec{1}}{24N^2}\right)
\ee
and
\be
\Tr(\rho^n_R)=\frac{2}{8^n}+\frac{2\cdot 9^n N^2}{(24)^nN^{2n}}=2^{1-3n}+\frac{2\,3^{n}}{2^{3n}N^{2(n-1)}}
\ee
This way the large $N$ Renyi entropy for $J=3$ becomes
\be
\Delta S^{(n)}_{R}=\frac{3n-1}{n-1}\log 2
\ee
hence we get a perfect agreement with \eqref{free}. Note again that $1/N$ corrections contain powers of $n-1$ hence they will contribute to von-Neumann entropy at $n=1$.

\end{appendix}

\end{document}